\documentclass[journal]{IEEEtran}
\usepackage{amsmath,amsfonts,amssymb,amsthm}
\usepackage{algorithmic}
\usepackage{stfloats}
\usepackage{algorithm}
\usepackage{array}
\usepackage{caption}
\usepackage[caption=false,font=normalsize,labelfont=sf,textfont=sf]{subfig}
\usepackage{textcomp}
\usepackage{url}
\usepackage{verbatim}
\usepackage{graphicx}
\usepackage{epstopdf}
\usepackage{cite}
\usepackage{float}
\usepackage{bm}
\usepackage{multirow}
\usepackage{microtype}
\usepackage{setspace}
\usepackage{arydshln}
\renewcommand{\arraystretch}{1.3}
\usepackage[hidelinks]{hyperref}
\makeatletter
\let\showhyphens\@undefined
\makeatother
\usepackage{orcidlink}
\newcommand{\mScale}[1]{\scalebox{0.92}{$\displaystyle #1$}}
\begin{document}
\title{Movable Antenna Assisted Flexible Beamforming for Integrated Sensing and Communication in Vehicular Networks}
\author{Luyang~Sun\textsuperscript{\orcidlink{0009-0009-0988-0087}},~\IEEEmembership{Graduate Student Member,~IEEE,} Zhiqing Wei\textsuperscript{\orcidlink{0000-0001-7940-2739}},~\IEEEmembership{Member,~IEEE,} Haotian Liu\textsuperscript{\orcidlink{0009-0000-0980-0440}},~\IEEEmembership{Graduate Student Member,~IEEE,} Kan Yu\textsuperscript{\orcidlink{0000-0003-0579-9985}},~\IEEEmembership{Member,~IEEE,} Zhendong Li\textsuperscript{\orcidlink{0000-0002-2928-193X}}, and Zhiyong Feng\textsuperscript{\orcidlink{0000-0001-5322-222X}},~\IEEEmembership{Senior Member,~IEEE}
\thanks{Luyang Sun, Zhiqing Wei, Haotian Liu, and Zhiyong Feng are with the School of Information and Communication Engineering, Beijing University of Posts and Telecommunications, Beijing
100876, China (e-mail: sly1105@bupt.edu.cn;
weizhiqing@bupt.edu.cn; haotian\_liu@bupt.edu.cn; fengzy@bupt.edu.cn).

Kan Yu is with the School of Computer Science and Engineering, Macau University of Science and Technology, Taipa, Macau 999078, China, and also with the Key Laboratory of Universal Wireless Communications, Ministry of Education, Beijing University of Posts and Telecommunications, Beijing 100876, China (e-mail: kanyu1108@126.com). 

Zhendong Li is with the Department of Electronic Engineering, Shanghai Jiao Tong University, Shanghai 200240, China, and also with the School of Information and Communication Engineering, Xi’an Jiaotong University, Xi’an 710049, China (e-mail: lizhendong@xjtu.edu.cn). (Corresponding authors: Zhiqing Wei, Zhiyong Feng)}}

\maketitle

\begin{abstract}
Integrated sensing and communication (ISAC) has been recognized as
a key technology in sixth-generation wireless networks,
and the additional spatial degrees of freedom obtained by movable antenna (MA)
technology can significantly improve the performance of ISAC systems.
This paper considers an ISAC-assisted vehicle-to-infrastructure (V2I) network, {\color{blue} where extended kalman filter-based prediction is combined with real-time optimization to jointly optimize transmit antenna positions and beamforming and power allocation vectors in dynamic environments. We propose two algorithms: a preprocessing-schur complement-projected gradient ascent algorithm for scenarios without sensing quality of service (QoS) constraints, which explores the potential range of sensing performance to provide reference and warm-starting for subsequent constrained optimization; and a heuristic reflective projected dynamic particle swarm optimization algorithm for sensing QoS-constrained scenarios, which achieves substantial performance gains under non-convex constraints with a small number of iterations. Simulation results demonstrate that these approaches enhance both the communication sum-rate and the lower of the Cramér-Rao lower bound of motion parameter estimation, validating the effectiveness of MA-assisted beamforming in dynamic V2I ISAC networks.}
\end{abstract}


\section{Introduction}
\subsection{Background and Motivations}
\IEEEPARstart{D}{riven} by the growing demand for high-precision localization,
environment-aware services,
and intelligent decision-making in the scenarios such as autonomous driving,
low-altitude economy, and industrial Internet of things,
the sixth-generation wireless network aims to
achieve a deep integration of the real physical world and the virtual digital world,
building a new world of intelligent interconnection and digital twin \cite{ref1}.
In this evolution, sensing and communication
systems are developing towards high frequency bands,
large antenna arrays, and miniaturization,
resulting in a gradual convergence of hardware architecture, channel characteristics,
and signal processing methods,
giving rise to integrated sensing and communication (ISAC) technology \cite{ref2}.
As a key technology to support new wireless services \cite{ref3}, 
ISAC is expected to use wireless communication signals to realize target detection, 
location, identification, imaging, etc., 
and then reconstruct the surrounding environment information.

With the rapid development of massive multiple input multiple output (mMIMO) technologies and millimeter wave (mmWave) technologies, 
ISAC has shown significant potential in communication and sensing performance \cite{ref4}. 
In addition, the sparsity of the mmWave mMIMO channel,
characterized by few non-line-of-sight (NLoS) components, 
is beneficial for vehicle positioning \cite{ref5}.
Vehicle-to-infrastructure (V2I) communication \cite{ref6,ref7},
as an indispensable component of the vehicle-to-everything network, 
facilitates information exchange between vehicles and roadside units (RSUs). 
This scenario imposes stringent requirements on both communication sum-rate and sensing accuracy, 
motivating extensive research on ISAC-enabled V2I systems that aim to simultaneously support high-rate communication and high-resolution positioning.

In particular, the mmWave-based ISAC system has recently attracted significant attention, 
offering a promising solution for high data rate and high sensing resolution in high-mobility V2I networks.
The system improves communication performance while ensuring sensing performance through beam tracking and prediction of motion parameters.
\cite{ref4} and \cite{ref8} both proposed adopting a cascaded scheme for channel prediction and beam alignment.
The principle of this design scheme is to achieve a trade-off between communication and sensing by
employing multiple antennas to steer signals in specific directions.
However, due to the fixed and discrete deployment of antennas, 
beamforming is strongly dependent on fixed channel conditions \cite{ref9}.
Even with advanced optimization techniques, the system can only attain a trade-off 
between communication sum-rate and sensing estimation accuracy under shared resources, without enabling their simultaneous improvement. 
The root of this limitation lies in the fact that once conventional fixed arrays are deployed, 
their positions and orientations cannot be altered, 
leading to severely constrained spatial degrees of freedom (DoFs).
With the rapid variations of channel conditions caused by high mobility, 
it becomes increasingly difficult for fixed arrays to simultaneously sustain communication and sensing performance in dynamic environments.
Therefore, it is urgent to introduce additional spatial DoF to achieve flexible regulation of the spatial response of the array, 
thereby breaking through the bottleneck of the existing physical structure.

Movable antenna (MA) technology has become a potential solution to overcome the above limitation \cite{ref10}. 
The MA is connected to the radio frequency chain through
a flexible cable and can be moved in a spatial region \cite{ref11}. 
In communication systems, compared with fixed-position antennas (FPAs),
MAs can be deployed to positions with more favorable channel conditions by 
moving within a designated region to actively
change the steering vector corresponding to different angles,
thereby flexibly adjusting the communication performance.

\subsection{Related Work}
\cite{ref12} verified that the MA-assisted communication system significantly 
improves the performance of diversity and spatial multiplexing.
\cite{ref13} improved the channel capacity by jointly optimizing 
the MA position in the transceiver and 
the covariance matrix of the transmitted signal under the MA-assisted MIMO system.
Moreover, \cite{ref14} proposed a new wireless sensing system 
equipped with MA array,
demonstrating that the angle estimation performance in wireless sensing 
is fundamentally determined by the geometry of the array. 
This conclusion revealed the potential for improving the sensing performance 
using an additional DoF with the same number of antennas.
\cite{ref15} further verified that the application of six-dimensional MA 
to the BS improves sensing performance within a given region. 

The effectiveness of MA in communication and sensing systems
is the basis for the research of MA technology in the ISAC system.
\cite{ref16} first studied the multi-user MIMO-ISAC system assisted by
the fluid antenna system (FAS) with sensing constraints.
An end-to-end learning framework for jointly optimizing the activation ports
and precoding design problem of a two-dimensional FAS is established using deep reinforcement learning,
which maximizes the sum-rate of MIMO downlink users
subject to sensing constraints.
Recently, MA technology has been shifting from discrete port selection
to continuous position optimization,
research on one-dimensional MA is in its early stages \cite{ref17}.
\cite{ref18} maximized communication sum-rate and sensing mutual information
by jointly beamforming and antenna position optimization.
\cite{ref19} maximized achievable data rate while meeting the sensing beam pattern gain requirements
by jointly optimizing the transmission information of the MA array,
sensing beamforming, and antenna position. 
\cite{ref20} maximized system capacity under the
radar sensing rate constraint by jointly optimizing the antenna positions of
the transceiver and transmit beamforming in the presence of clutter.
\cite{ref21} maximized the weighted-sum of communication capacity and sensing mutual information
by jointly optimizing beamforming and antenna positions at the transceiver.
In addition, \cite{ref22} minimized the sensing Cramér-Rao lower bound (CRLB) to improve parameter estimation performance.
However, most existing studies are based on static targets,
and the evaluation of sensing performance mainly focuses on the CRLB of angle estimation, 
with less attention paid to actual traffic scenarios with dynamic characteristics.

\begin{table*}[!t]
{\color{blue}
\caption{\underline{\textbf{This table is newly added}}{\color{blue} Comparison of the main contributions of this paper with other related papers.}}
\footnotesize
\setlength{\tabcolsep}{2pt}
\label{tab1}
\scalebox{0.8}{
\begin{tabular}{|c|c|c|c|c|c|c|c|}
\hline
  \textbf{Reference}& \textbf{Move Model} & \textbf{Application scenario}&\textbf{Dynamic / Static}&   \textbf{Main contributions}  \\
\hline
 \cite{ref16} & Discrete & MU-MIMO&Static& jointly optimize antenna port locations and precoding design to improve overall communication rate  \\
\hline
\cite{ref18}&Continuous  & MU-SIMO& Static&   communication rate
 and sensing mutual information maximization \\ 
\hline
\cite{ref19} &Continuous   &MU-SIMO &Static&  improve the throughput capacity and meet the requirement of
 the sensing beampattern threshold  \\
\hline
\cite{ref20} &Continuous  &MU-MISO  &Static&  optimize MA positions and beamforming to maximize capacity under radar sensing rate constraints \\
\hline
\cite{ref21} & Continuous  & MU-MISO&Static& the self-interference channel is modeled as a function of the antenna position vectors under the near-field channel condition  \\
\hline
\cite{ref22} & Continuous  & MU-SIMO&Static& minimize the Cramér-Rao bound (CRB) for estimating the target's angle while guaranteeing communication performance\\
\hline
this paper & Continuous  & MU-MISO & Dynamic&  combining MA with beam prediction to maximize communication sum-rate under sensing constraints in dynamic scenarios\\
\hline
\end{tabular}}}
\end{table*}

\subsection{Our Contributions}
{\color{blue}Based on the analysis presented above,} this paper will further explore the effectiveness of MA-assisted V2I network flexible beamforming.
The geometric relationship of vehicles relative to the RSU at any time can be characterized by parameters such as direction of departure (DoD), distance, and velocity. 
These geometric parameters remain unchanged within tens of milliseconds.
{\color{blue}
Therefore, we propose a flexible beamforming framework for dynamic ISAC tasks, which integrates a prediction stage with a real-time adjustment stage. 
In the prediction stage, the system jointly optimizes MA positions, the transmit beam phases, and the power allocation to obtain the optimal transmit antenna configuration for the next time slot. 
In the subsequent real-time adjustment stage, only the previously optimized transmit antenna positions are used, while the power and beam phases are adjusted based on the current estimated channel to accurately align the beam with the target.}
The main contributions of this paper are summarized as follows.

\begin{itemize}
    \item \textbf{\textit{Comprehensive sensing performance metric:}} 
    We extend the conventional CRLB-based metric (mainly limited to angle estimation) 
    to a more comprehensive {\color{blue}the lower of Cramér-Rao lower bound (LCRLB)} that jointly accounts for angle, distance, and velocity.
    {\color{blue}
    Then, by incorporating vehicle motion models into dynamic V2I scenarios, we transform the LCRLB into the lower of posterior CRLB (LPCRLB) to characterize sensing performance in time-varying environments.
    We further derive the functional relationship between the motion parameter LPCRLB and the positions of transmit and/or receive MA, revealing how the sensing performance depends on antenna deployment.}
    This provides theoretical support for subsequent MA position optimization 
    and the trade-off between communication and sensing performance.
    {\color{blue}
    \item \textbf{\textit{Algorithmic framework:}} 
    We propose an extended kalman filter (EKF)-enhanced two-regime optimization framework for dynamic ISAC systems, combining an exploration mode without sensing quality of service (QoS) constrain to reveal intrinsic channel and antenna effects, and a QoS-aware dynamic mode with warm-start pre-optimization and real-time refinement to ensure sensing accuracy under mobility.
    This hybrid approach significantly improves optimization stability and tracking performance compared with conventional methods.
    \item \textbf{\textit{Algorithmic Design:}} 
    Building upon the proposed framework, we develop two specific algorithms tailored to the two regimes. 
    For scenarios without sensing QoS constraints, a preprocessing–Schur complement–projected gradient ascent (PRE-SC-PGA) algorithm efficiently maximizes the weighted-sum with low complexity and guaranteed convergence. 
    For QoS-constrained scenarios, a ``heuristic'' reflective projected dynamic particle swarm optimization (RPDPSO) algorithm optimizes the transmit antenna positions via a small number of external iterations, achieving substantial performance gains for non-convex constraints with minimal computational cost.
    \item \textbf{\textit{Performance verification:}}
   We validate the effectiveness of the proposed framework and algorithms through extensive simulations in a V2I system. 
   The results demonstrate that the MA-assisted schemes can significantly improve communication and sensing performance, highlighting the potential of MA to overcome the inherent limitations of fixed arrays. 
    This step closes the loop by connecting the framework and algorithm design to practical system benefits.}
\end{itemize}

{\color{blue} The comparison of the main contributions between the related works and this paper is summarized in \textbf{Table I.}
} 
The remainder of this work is organized as follows. 
The system model of the MA-assisted ISAC based on the V2I network is introduced in Section II {\color{blue} and the expression of LCRLBs is derived. }
Section III {\color{blue}constructs a weighted-sum maximization problem and a communication performance maximization problem under sensing constraints for scenarios with and without sensing QoS constraints.} 
To address this problem, {\color{blue}the PRE-SC-PGA and RPDPSO algorithms are proposed in Sections IV and Section V, respectively.}
Section  {\color{blue}VI} simulates to verify the effectiveness and superiority of the proposed {\color{blue} schemes}. 
Section  {\color{blue}VII} concludes this paper.

\textit{Notations}: The bold uppercase letter,
bold lowercase letter and the normal font represent the matrix, vector, and scalar, respectively. 
$\mathbf{W}\succeq 0$ indicates that $\mathbf{W}$ is a positive semidefinite matrix.
$\left ( \cdot  \right ) ^{\ast }, \left ( \cdot  \right ) ^{T}$ and $ \left ( \cdot  \right ) ^{H}$ denote complex conjugate, 
transpose, and conjugate transpose, respectively.
$\mathrm{det}\left ( \cdot  \right ) $ is the matrix determinant.
$\mathrm{Tr}\left ( \cdot  \right ) $  and $\mathrm{rank}\left ( \cdot  \right ) $ denote the trace and rank of a square matrix, respectively.
$\mathbb{C}$ and $\mathbb{R}$ represent the sets of complex numbers and real numbers, respectively.
$\mathcal{CN}(\mu,\mathbf{\Sigma }  )$ denotes the circularly symmetric complex Gaussian (CSCG) distribution,
where $\mu$ and $\mathbf{\Sigma } $ are the mean vector and the covariance matrix, respectively.
The notation $\oplus$ is Khatri-Rao product {\color{blue} and $\odot$ is Hadamard product.}
$\mathbb{E}$ denotes the expectation operator.
$\mathbf{I}_{M_{rx}Q}$ denotes the $M_{rx}Q$-dimensional identity matrix. 
{\color{blue}
$\delta(\cdot)$ is a counting function.
Other important parameters are listed in Table II.}

\begin{table*}[!t]
\centering
{\color{blue}
\caption{\underline{\textbf{This table is newly added}}{\color{blue} Parameter Description}}
\footnotesize
\setlength{\tabcolsep}{4pt}
\renewcommand{\arraystretch}{1.08}
\label{tab2}
\scalebox{0.72}{
\begin{tabular}{|c|c|c|c|c|c|}
\hline
\textbf{Parameter} & \textbf{Description} & \textbf{Parameter} & \textbf{Description}& \textbf{Parameter} & \textbf{Description}\\
\hline
$M_{tx},M_{rx}$ & Number of transmit/receive antennas & $D_{\min}^{tx},D_{\max}^{tx},D_{\min}^{rx},D_{\max}^{rx}$ & Feasible range of antenna positions&$\imath$ & Penalty parameter\\
\hline
$\mathbf{p}_{tx},\mathbf{p}_{rx}$ & Transmit/receive antenna position vectors & $\mathcal{N}_{k}$ & Subcarrier set for user $k$ &$w_{\min},w_{\max}$& Inertial weight bounds\\
\hline
$K$ & Number of vehicles & $\mathbf{a}(\mathbf{p}_{tx},\theta_k),\ \mathbf{b}(\mathbf{p}_{rx},\theta_k)$ & Steering vectors&$c_1,c_2$ & Individual and global learning factors\\
\hline
$N$,$N_{\bar{p}}$ & Number of subcarriers, Number of particles&$\zeta$ & Path-loss exponent&$\varsigma_{k}^{\theta},\varsigma_{k}^{d} \varsigma_{k}^{\nu}$ & Known sensing LPCRLBs\\
\hline
$Q$ & Number of OFDM symbols & $\varphi$ & Phase shift&$\mathbf{u}_{pbest,\bar{p}},\ \mathbf{u}_{gbest,\bar{p}}$ & Personal and global best positions\\
\hline
$\Delta f$ & Subcarrier spacing &$\eta_0,\eta_1,\mathbf{n}$ & Power spectral densities, Gaussian noise terms& $P_{act}$ & Active particles\\
\hline
$T_s, T_e, T_{cp}$ & Symbol duration, Signal length, Cyclic prefix & $\mathbf{w}$, $P_T$,& Beamforming vector, Transmit power & $\mathbf{u}_{\bar{p}},\ \mathbf{v}_{\bar{p}}$ & particle position and velocity\\
\hline
$\mathbf{p}$,$\mathbf{p}_{n}$ & Transmit power vector& $s_{f}^{1},s_{f}^{2},s_{f}^{3}$ &Scaling factor& $\mathcal{P},\ \mathcal{S}$ & Penalty function, sum-rate function\\
\hline
$p(\cdot)$ & Conditional PDF & $\aleph$, $\rho$ & Quantization factor, Weighting factor& $\mathrm{Iter}$ & Number of iterations\\
\hline
$D_{sp},D_{tr}$ & Minimum distance between adjacent antennas & $\delta_1,\delta_2$ & Update step lengths& $\mathbf{g},\ \mathbf{h}$ & Element vector, communication channel\\
\hline
$\beta$ & Attenuation coefficient & $\mu$,$\tau$ & Doppler frequency, Time delay& $\boldsymbol{g}(\cdot),\boldsymbol{h}(\cdot)$ & State and measurement functions\\
\hline
$D_{\max}$ & Antenna feasible region & $\Delta T,\Delta\nu,\Delta d$ & Time duration, velocity and distance increment& $\mathbf{L}$ & LPCRLB diagonal matrix\\
\hline
$\theta,d,\nu$ & Angle, Distance, Velocity &$\alpha_0$,$\alpha$  & Path-loss constant,Path-loss coefficient& $\mathbf{u},\ \boldsymbol{\zeta}$ & Observation and motion parameter vector\\
\hline
\end{tabular}}}
\end{table*}

\section{System Model}
As illustrated in Fig. 1, we consider an MA-ISAC-assisted V2I monostatic scenario{\color{blue},  where all vehicles are always parallel to the road and
the direction of their velocity vectors remains almost constant.}  
The RSU simultaneously serves $K$ single-antenna mobile vehicles and
receives the reflected signal echoes for sensing.
In the RSU, a massive MIMO linear MA array of mmWave is equipped with $M_{tx}$ transmit antennas and $M_{rx}$ receive antennas,
where the position of the MA can be flexibly adjusted in a one-dimensional line segment of a given length. 

\begin{figure}[!t]
\centering
\includegraphics[width=2in]{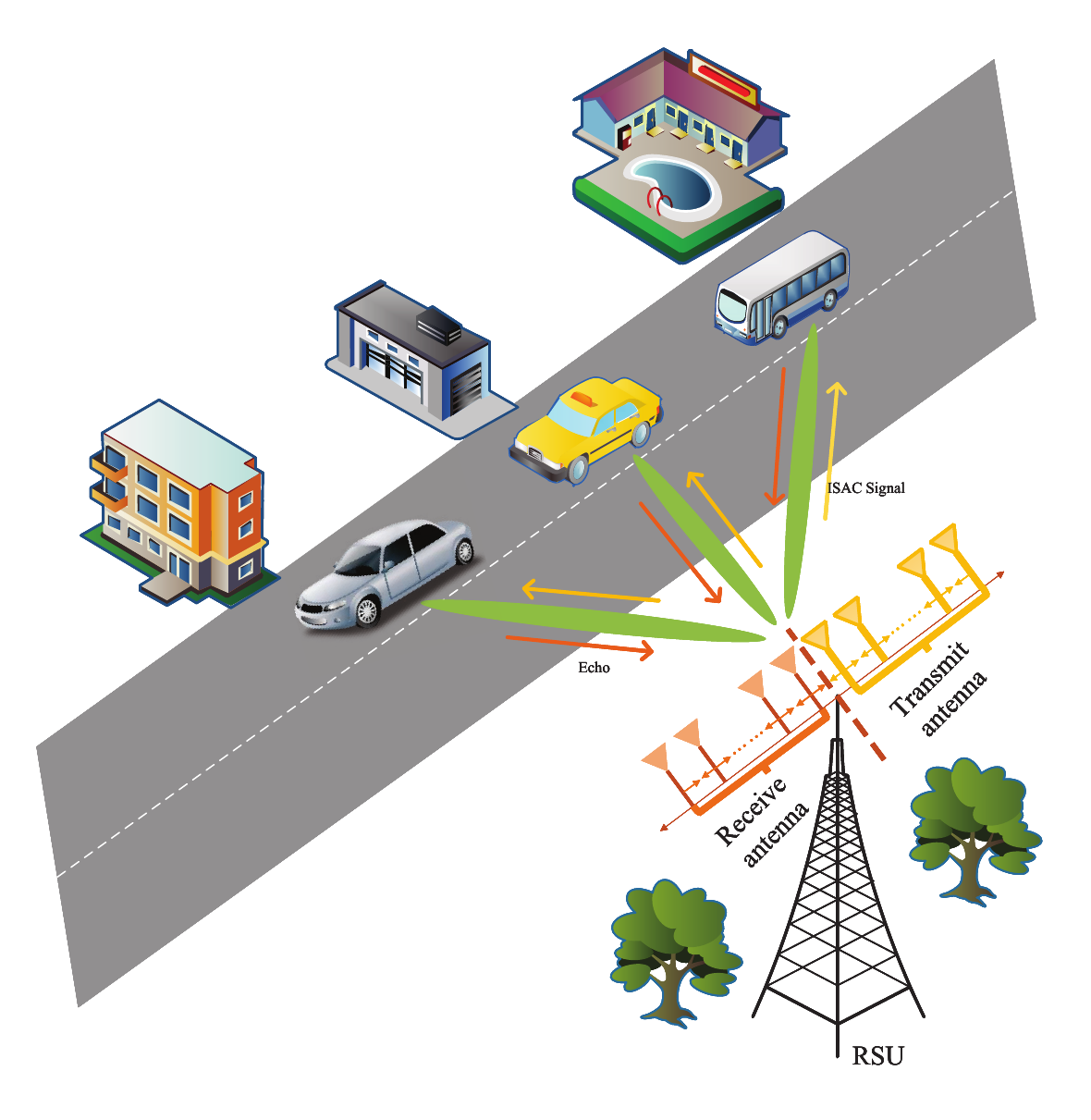}
\captionsetup{justification=raggedright,singlelinecheck=false}
\caption{\underline{\textbf{This figure is updated}}MA-ISAC-assisted V2I system.}
\label{fig1}
\end{figure}

\subsection{Signal Model}
To focus on the effects of antenna positions and beamforming on communication and sensing performance, we assume a physical separation to suppress the self-interference caused by direct signal leakage from the transmit antennas to the receive antennas \cite{ref23} and do not consider additional interference sources.
Let $p_{tx,l}\in\left[D_{min}^{tx}, D_{max}^{tx}\right]$ and $p_{rx,ll}\in\left[D_{min}^{rx}, D_{max}^{rx}\right]$ represent the position of the $l$-th transmit MA and the $ll$-th receive MA, respectively. 
Then, the transmit antenna position vector (APV) can be expressed as $\mathbf{p}_{tx}$ $= \left [p_{tx,1},p_{tx,2},...,p_{tx,M_{tx}} \right ]^{T}\in\mathbb{R}^{M_{tx}\times1}$, and the receive APV can be expressed as $\mathbf{p}_{rx}$ $=\left [p_{rx,1},p_{rx,2},...,p_{rx,M_{rx}} \right ]^{T}\in\mathbb{R}^{M_{rx}\times1}$,
where {\color{blue}$D_{min}^{tx}\le p_{tx,1}<p_{tx,2}<\dots< p_{tx,M_{tx}}\le D_{max}^{tx}$, $D_{min}^{rx}\le p_{rx,1}<\dots < p_{rx,M_{rx}}\le D_{max}^{rx}$ \cite{ref22}}.

To facilitate the transmission of information bits to $K$ distinct vehicles, 
the available subcarriers are divided into disjoint subsets $\mathcal{N}_{k}$, 
each of which is exclusively allocated to a single vehicle \cite{ref24}. 
In particular, we assume that the sub-carriers are uniformly divided into three consecutive groups in advance, 
intended for use by three vehicles.
Without loss of generality, the transmitted baseband orthogonal frequency division multiple (OFDM) signal is divided into $Q$ blocks, with $N$ orthogonal subcarriers in each block. 
Let the cyclic prefix length be $T_{cp}$, the frequency interval between adjacent subcarriers be $\bigtriangleup f = 1/T_{e}$, 
then the duration of each block is $T_{s} = T_{e}+T_{cp}$. 
The baseband signal transmitted by the RSU in the $q$-th data block can be expressed as \cite{ref25}
\begin{equation}
\label{deqn_ex1a}
\scalebox{0.97}{$
\mathbf{\tilde{s}}_{q}(t)={\textstyle \sum\nolimits_{n=0}^{N-1}}\mathbf {w}_{n}c_{n}^{q}e^{j2\pi n\bigtriangleup f\left(t-qT_{s}\right)} rect[\frac{t-qT_{s}}{T_{s}}], n\in\mathcal{N}_{k},$}
\end{equation}
where $\bf w$$_{n}\in \mathbb{C}^{M_{tx}\times1}$ is the beamforming vector of the $n$-th subcarrier, $c_{n}^{q}$ is the transmission symbol of the $n$-th subcarrier in the $q$-th data block. 
$rect [\frac{t}{T_{s}}]$ is a rectangular function that is equal to 1 when $T_{s}$ and 0 otherwise. 
Therefore, the signals transmitted over the $Q$ data blocks are $\mathbf{\tilde{s}}(t)={\sum_{q=1}^{Q}} \mathbf{\tilde{s}}_{q}(t)$.

In the mmWave communication system, a LoS channel is generally adopted \cite{ref26}. 
In fact, the presence of obstacles can hinder the execution of communication and sensing tasks, 
and echoes from NLoS channels can mislead the expected position of the target. 
For the sake of convenience, we will leave the impact of these factors to future work and will only consider the LoS channel in this paper.
In addition, because the distance between the vehicle and the RSU is much larger than the size of the antenna movement region, 
the far-field channel model is considered.
 
let the transmit steering angle of the transmitter-vehicle be $\theta_{k}\in \left[0,\pi\right]$,
the transmit steering vector of MA is a function of $\mathbf{p}_{tx}$ and $\theta_{k}$, i.e., \cite{ref18}
\begin{equation}
\label{deqn_ex2a}
\mathbf{a}(\mathbf{p}_{tx}, \theta_{k})=[e^\frac{j2\pi p_{tx,1} cos\theta_{k}}{\lambda},\cdots,e^\frac{j2\pi p_{tx,M_{tx}} cos\theta_{k}}{\lambda}]^{T}.
\end{equation}
Similarly, {\color{blue}the receive steering vector can be written as \cite{ref16}
\begin{equation}
\label{deqn_ex3a}
\mathbf{b}(\mathbf{p}_{rx},\theta_{k})=[e^\frac{j2\pi p_{rx,1}cos\theta_{k}}{\lambda},\cdots,e^\frac{j2\pi p_{rx,M_{rx}} cos\theta_{k}}{\lambda}]^{T}.
\end{equation}}

\subsection{Sensing Model}
In this case, let $\nu_{k}$ and $d_{k}$ represent the velocity of the $k$-th vehicle and the length of the two propagation paths, respectively. 
At this time, the received signal echoes at the RSU from vehicles can be expressed as \cite{ref4}
\begin{equation}
\label{deqn_ex4a}
\scalebox{0.84}{$
\begin{aligned}
\mathbf{r}(t)=\sum_{k=1}^{K}\sum_{q=1}^{Q}\beta_{k}e^{j2\pi\mu_{k}\left(q-1\right)T_{s}}\mathbf{b}(\mathbf{p}_{rx},{\color{blue}\theta_{k}})\mathbf{a}^{H}(\mathbf{p}_{tx},\theta_{k})\mathbf{\tilde{s}}_{q}(t-
\tau_{k})+\mathbf{z}_{k}^{q}(t),
\end{aligned}$}
\end{equation}
where $\beta_{k}$ is the attenuation coefficient including two propagation losses and the reflection coefficient, $\mu_{k}=\frac{{\color{blue}2cos\theta_{k}}\nu_{k}}{\lambda}$ and $\tau_{k}=\frac{{\color{blue}2}d_{k}}{c}$ are the Doppler frequency and the time delay with respect to the $k$-th vehicle, respectively. 
$\mathbf{z}_{k}^{q}(t)\in\mathbb{C}^{M_{rx}\times1}$ denotes the CSCG noise vector {\color{blue}with power spectral density $\eta_{1}$} in the RSU.

After $\nu_{k}$ and $d_{k}$ are estimated by the traditional matched-filtering method,
the received echoes in the $n$-th subcarrier of the $q$-th block can be rewritten as \cite{ref26}
\begin{equation}
\label{deqn_ex5a}
\scalebox{0.8}{$
\begin{aligned}
\tilde{\mathbf{y}}_{n}^{q}&=\int_{0}^{{\color{blue}T_{e}}}\mathbf{r}(t)c_{n}^{*\left(q\right)}e^{-j2\pi n\triangle f\left(t-qT_{s}\right)}dt\\&=\sum_{k=1}^{K}\beta_{k}{\color{blue}T_{e}p_{n}}\mathbf{b}(\mathbf{p}_{rx},{\color{blue}\theta_{k}})\mathbf{a}^{H}(\mathbf{p}_{tx},\theta_{k})\mathbf{w}_{n}e^{-j2\pi n\triangle f\tau_{k}}e^{j2\pi\mu_{k}(q-1)T_{s}}+\tilde{\mathbf{z}}_{n}^{q},
\end{aligned}$}
\end{equation}
where $\tilde{\mathbf{z}}_{n}^{q}=\int_{0}^{{\color{blue}T_{e}}}\mathbf{z}_{k}^{q}(t)c_{n}^{*\left(q\right)}e^{-j2\pi n\triangle f\left(t-qT_{s}\right)}dt$ is a complex Gaussian vector with zero mean and covariance {\color{blue}$ p_{n}\eta_{1}T_{e}\mathbf{I}_{M_{rx}}$}. 
$\gamma_{k}=\beta_{k}{\color{blue}T_{e}}$ and
{\color{blue}$\mathbf{C}=[\mathbf{c}(\mu_{1}),\mathbf{c}(\mu_{2}),\cdots,\mathbf{c}(\mu_{K})]$}, where ${\color{blue}\mathbf{c}(\mu_{k})}=\left[\gamma_{k},\gamma_{k}e^{j2\pi \mu_{k}T_{s}},\cdots,\gamma_{k}e^{j2\pi \mu_{k}\left(Q-1\right)T_{s}}\right]^{T}$. 
Based on {\color{blue}(\ref{deqn_ex5a})}, the received echoes of the $q$-th block can be modeled in the form of a matrix, i.e., \cite{ref27}
\begin{equation}
\label{deqn_ex6a}
\scalebox{0.83}{$
\begin{aligned}
\tilde{\mathbf{Y}}^{q}=\sum_{k=1}^{K}\gamma_{k}\mathbf{b}(\mathbf{p}_{rx},{\color{blue}\theta_{k}})e^{j2\pi\mu_{k}(q-1)T_{s}}{\color{blue}\boldsymbol{\chi}\left(\tau_{k}\right)}+\tilde{\mathbf{Z}}_{q}=\mathbf{B}\mathrm{diag}(\mathbf{C}_{q})\mathbf{\Gamma}^{T}+\tilde{\mathbf{Z}}^{q},
\end{aligned}$}
\end{equation}
where $\mathbf{B}=[\mathbf{b}(\mathbf{p}_{rx},{\color{blue}\theta} _{1}),\cdots,\mathbf{b}(\mathbf{p}_{rx},{\color{blue}\theta} _{K})]$, $\mathbf{C}_{q}$ denotes the $q$-th row data of $\mathbf{C}$. 
$\mathbf{\Gamma}=\left[\boldsymbol{\chi}(\tau_{1}),\cdots,\boldsymbol{\chi}(\tau_{K}\right)]$, $\boldsymbol{\chi}(\tau_{k})=\mathbf{a}^{H}(\mathbf{p}_{tx},\\\theta_{k})[{\color{blue}p_{1}}\mathbf{w}_{1},
\cdots,{\color{blue}p_{N}}\mathbf{w}_{N}e^{-j2\pi(N-1)\Delta f\tau_{k}}]$, $\tilde{\mathbf{Z}}^{q}=\left[\tilde{\mathbf{z}}_{1}^{q},\cdots,\tilde{\mathbf{z}}_{N}^{q}\right]$.  
In addition, the transmitted signals over the $Q$ data blocks can be rewritten as
$\tilde{\mathbf{Y}}=\left(\mathbf{B}\oplus \mathbf{C}\right)\mathbf{\Gamma}^{T}+\tilde{\mathbf{Z}}=\tilde{\mathbf{X}}+\tilde{\mathbf{Z}}$, 
where $\tilde{\mathbf{Z}}=\big[\tilde{\mathbf{Z}}^{1},\tilde{\mathbf{Z}}^{2},\cdots,\tilde{\mathbf{Z}}^{Q}\big]^{T}$.

\begingroup
\color{blue}
For sensing performance, our aim is to evaluate its accuracy in position estimation, 
which is usually evaluated by CRLB.
Initially, by stacking the columns of $\tilde{\mathbf{Y}}$ vertically, we can obtain $\tilde{\mathbf{y}}=\tilde{\mathbf{x}}+\tilde{\mathbf{z}}$, 
where $\tilde{\mathbf{y}}\sim\mathcal{CN}\left(\tilde{\mathbf{x}},\mathbf{\Sigma}\right)$ with $\mathbf{\Sigma}=\mathrm{diag}{\color{red}(p_{1}\eta_{1}T_{e}\mathbf{I}_{M_{rx}Q},p_{2}\eta_{1}T_{e}\mathbf{I}_{M_{rx}Q},\cdots,p_{N}\eta_{1}T_{e}\mathbf{I}_{M_{rx}Q})}$. 
Then, let $\phi_{k}=cos(\theta_{k})$
and define the observation vector and the motion parameter vector to be estimated as $\mathbf{u} = [ \mathbf{u} _{1}^{T},\mathbf{u} _{2}^{T},\cdots , \mathbf{u} _{K}^{T}  ]^{T} $ and $\bm{\zeta} =\left [ \bm{\zeta}_{1}^{T},\bm{\zeta} _{2}^{T},\cdots , \bm{\zeta}_{K}^{T} \right ]^{T} $, respectively, where {\color{red}$\mathbf{u} _{k}=\left [\phi_{k},\tau_{k},\mu_{k} \right ] $ and $\bm{\zeta} _{k}=\left [\theta_{k},d_{k},\nu_{k} \right ] $}. 
Given $\mathbf{u}$, the conditional probability density function of $\tilde{\mathbf{y}}$ can be expressed as \cite{ref26}
\begin{equation}
\label{deqn_ex7a}
p\left ( \tilde{\mathbf{y} } |\mathbf{u}  \right ) =\frac{1}{\pi^{M_{rx}QN}\mathrm{det}(\mathbf{\Sigma})  }e^{-\left(\tilde{y}-\tilde{x} ^{H}\right)\mathbf{\Sigma}  ^{-1} \left(\tilde{y}-\tilde{x}\right)}.
\end{equation}
According to CRLB theory \cite{ref28}, the Fisher information matrix of $\mathbf{u}$ is given by
\begin{equation}
\label{deqn_ex8a}
\mathbf{J}( \mathbf{u})=\left ( \partial \tilde{\mathbf{x} } /\partial \mathbf{u} \right ) ^{H}\mathbf{\Sigma } ^{-1}\left (\partial \tilde{\mathbf{x} } /\partial \mathbf{u} \right ).
\end{equation}
Given $\forall n,q$, 
let $\left[\tilde{\mathbf{x}}\right]_{1+\left(q-1\right)M_{rx}}^{qM_{rx}}$ denote the $1+\left(q-1\right)M_{rx}$-th to $qM_{rx}$-th row of $\tilde{\mathbf{x}}$ and its partial derivative 
with respect to the $i$-th motion parameter of the $k$-th vehicle is expressed as $\boldsymbol{\varpi}_{nq}^{ik}$,
we have 
\begin{equation}
\label{deqn_ex9a}
\scalebox{0.9}{$
\begin{aligned}
\boldsymbol{\varpi}_{nq}^{1k} &=\gamma_{k}{\color{red}p_{n}(\bar{\Lambda}_{rx}\mathbf{b}( \mathbf{p}_{rx},\theta_{k})\mathbf{a}^{H} ( \mathbf{p} _{tx},\theta_{k} )+\mathbf{b}( \mathbf{p}_{rx},\theta_{k} )\mathbf{a}^{H} ( \mathbf{p} _{tx},\theta_{k} )}\\
&{\color{red}\times\mathbf{\bar{\Lambda}}_{tx})}\mathbf{w} _{n}e^{-j2\pi n\Delta f\tau_{k}} e^{j2\pi\mu_{k}\left (q-1  \right )T_{s}},\\
\boldsymbol{\varpi}_{nq}^{2k} &=-j2\pi n\Delta f\gamma_{k}{\color{red}p_{n}}\mathbf{b}( \mathbf{p} _{rx},{\color{red}\theta} _{k})\mathbf{a} ^{H}( \mathbf{p} _{tx},{\color{red}\theta}  _{k})\mathbf{w} _{n}e^{-j2\pi n\Delta  f\tau_{k}}\\
&\times e^{j2\pi\mu_{k}(q-1)T_{s}},\\
\boldsymbol{\varpi}_{nq}^{3k} &=j2\pi (q-1)T_{s}\gamma_{k}{\color{red}p_{n}}\mathbf{b}( \mathbf{p} _{rx},{\color{red}\theta} _{k})\mathbf{a}^{H} ( \mathbf{p} _{tx},{\color{red}\theta}  _{k})\mathbf{w}_{n}e^{-j2\pi n\Delta  f\tau_{k}}\\
&\times  e^{j2\pi\mu_{k}\left (q-1\right )T_{s}},
\end{aligned}$}
\end{equation}
where $\bar{\mathbf{\Lambda}}_{tx}=\frac{-j2\pi}{\lambda}\mathrm{diag}(\mathbf{p}_{tx})$ and $\bar{\mathbf{\Lambda}}_{rx}=\frac{j2\pi}{\lambda}\mathrm{diag}(\mathbf{p}_{rx})$.
Based on $\mathbf{J}(\mathbf{u})$, applying the chain rule \cite{ref26},
we can further obtain 
\begin{equation}
\label{deqn_ex10a}
\scalebox{0.85}{$
        \mathbf{J}(\bm{\zeta})=\mathbf{Q}\mathbf{J}(\mathbf{u})\mathbf{Q}^{T}=\left [ \begin{matrix}
  \boldsymbol{\mathcal{Q}}_{11} &\boldsymbol{\mathcal{Q}}_{12}  &\cdots  &\boldsymbol{\mathcal{Q}}_{1K} \\
\boldsymbol{\mathcal{Q}}_{21}&\boldsymbol{\mathcal{Q}}_{22}  & \cdots &\boldsymbol{\mathcal{Q}}_{2K} \\
  \vdots&\vdots   &\ddots  &\vdots \\
 \boldsymbol{\mathcal{Q}}_{K1} &\boldsymbol{\mathcal{Q}}_{K2}  &\cdots   &\boldsymbol{\mathcal{Q}}_{KK}
\end{matrix} \right ],$}
\end{equation}
where $\mathbf{Q}=\mathrm{diag}(\left[\mathbf{Q}_{kk}\right]_{k=1}^{K})$ and 
{\color{red}
\begin{equation}
    \begin{aligned}
    \label{deqn_ex11a}
    \mathbf{Q}_{kk}=\left [\begin{matrix}
  -sin\theta_{k}& 0 &  -\frac{2\nu_{k} sin\theta_{k}}{\lambda}\\
  0&  \frac{2}{c}&0 \\
 0& 0 &\frac{2cos\theta_{k}}{\lambda}
\end{matrix}  \right ]. 
 \end{aligned}
\end{equation}
Therefore, the principal diagonal block of $\mathbf{J}(\bm{\zeta})$ can be rewritten as
\begin{equation}
    \label{deqn_ex12a}
\scalebox{0.71}{$
\begin{aligned}
\bm{\mathcal{Q}}_{kk}=\left [ 
\begin{matrix}
 sin^{2}\theta_{k} \mathbf{p}^{T}\mathbf{g}_{kk}^{11} +\frac{4\nu_{k}sin^{2}\theta_{k}}{\lambda}(\mathbf{p}^{T}\mathbf{g}_{kk}^{31} +\frac{\nu_{k}}{\lambda}\mathbf{p}^{T}\mathbf{g}_{kk}^{33})& \bm{\mathcal{Q}}_{kk}^{21} &\bm{\mathcal{Q}}_{kk}^{31} \\
 -\frac{2sin\theta_{k}}{c}(\mathbf{p}^{T}\mathbf{g}_{kk}^{21} + \frac{2\nu_{k}}{\lambda}\mathbf{p}^{T}\mathbf{g}_{kk}^{23})& \frac{4}{c^{2}}\mathbf{p}^{T}\mathbf{g}_{kk}^{22} & \frac{4cos\theta_{k}}{c\lambda}\mathbf{p}^{T}\mathbf{g}_{kk}^{23} \\
 -\frac{2sin\theta_{k}cos\theta_{k}}{\lambda}(\mathbf{p}^{T}\mathbf{g}_{kk}^{31} +\frac{2\nu_{k}}{\lambda}\mathbf{p}^{T}\mathbf{g}_{kk}^{33})  &  \bm{\mathcal{Q}}_{kk}^{23}  &\frac{4cos^{2}\theta_{k}}{\lambda^{2}}\mathbf{p}^{T}\mathbf{g}_{kk}^{33}
\end{matrix} \right ],    
\end{aligned} $}
\end{equation}
where $\bm{\mathcal{Q}}_{kk}^{ij}$ is the $(i,j)$-th element of $\bm{\mathcal{Q}}_{kk}$.

\textit{Proof:} See appendix A.

According to the property $[\bm{\mathrm{J}}^{-1}(\boldsymbol{\zeta})]_{i+3(k-1),i+3(k-1)}\ge[\bm{\mathcal{Q}}_{kk}]_{ii}^{-1}, i =1,2,3$ in \cite{ref28}, we can approximately obtain the LCRLBs, which follows
\begin{equation}
\scalebox{0.9}{$
\begin{aligned}
\label{deqn_ex13a}
\mathrm{LCRLB}_{\theta_{k}}^{-1}& =  sin^{2}\theta_{k} \mathbf{p}^{T}\mathbf{g}_{kk}^{11} +\frac{4\nu_{k}sin^{2}\theta_{k}}{\lambda}(\mathbf{p}^{T}\mathbf{g}_{kk}^{31} +\frac{\nu_{k}}{\lambda}\mathbf{p}^{T}\mathbf{g}_{kk}^{33}),\\
\mathrm{LCRLB}_{d_{k}}^{-1} &= \frac{4}{c^{2}}(\mathbf{p}^{T}\mathbf{g}_{kk}^{22}-\frac{(\mathbf{p}^{T}\mathbf{g}_{kk}^{23}) ^{2}}{\mathbf{p}^{T}\mathbf{g}_{kk}^{33}}),\\
\mathrm{LCRLB}_{\nu_{k}}^{-1}& =\frac{4cos^{2}\theta_{k}}{\lambda^{2}}(\mathbf{p}^{T}\mathbf{g}_{kk}^{33}-\frac{(\mathbf{p}^{T}\mathbf{g}_{kk}^{23} )^{2}}{\mathbf{p}^{T}\mathbf{g}_{kk}^{22}}).
\end{aligned}$}
\end{equation}
Evidently, the LCRLBs are closely related to the vector $\mathbf{g}_{kk}$. 
As illustrated in the Fig.\ref{fig2}, increasing the number of receive antennas, subcarriers, and OFDM symbols consistently improves the sensing LCRLBs performance, indicating enhanced estimation accuracy due to the increased amount of available observation information.
The LCRLB of DoD remains unchanged because its estimation accuracy does not depend on the number of subcarriers.}

\begin{figure}[t]
\includegraphics[width=3.5in]{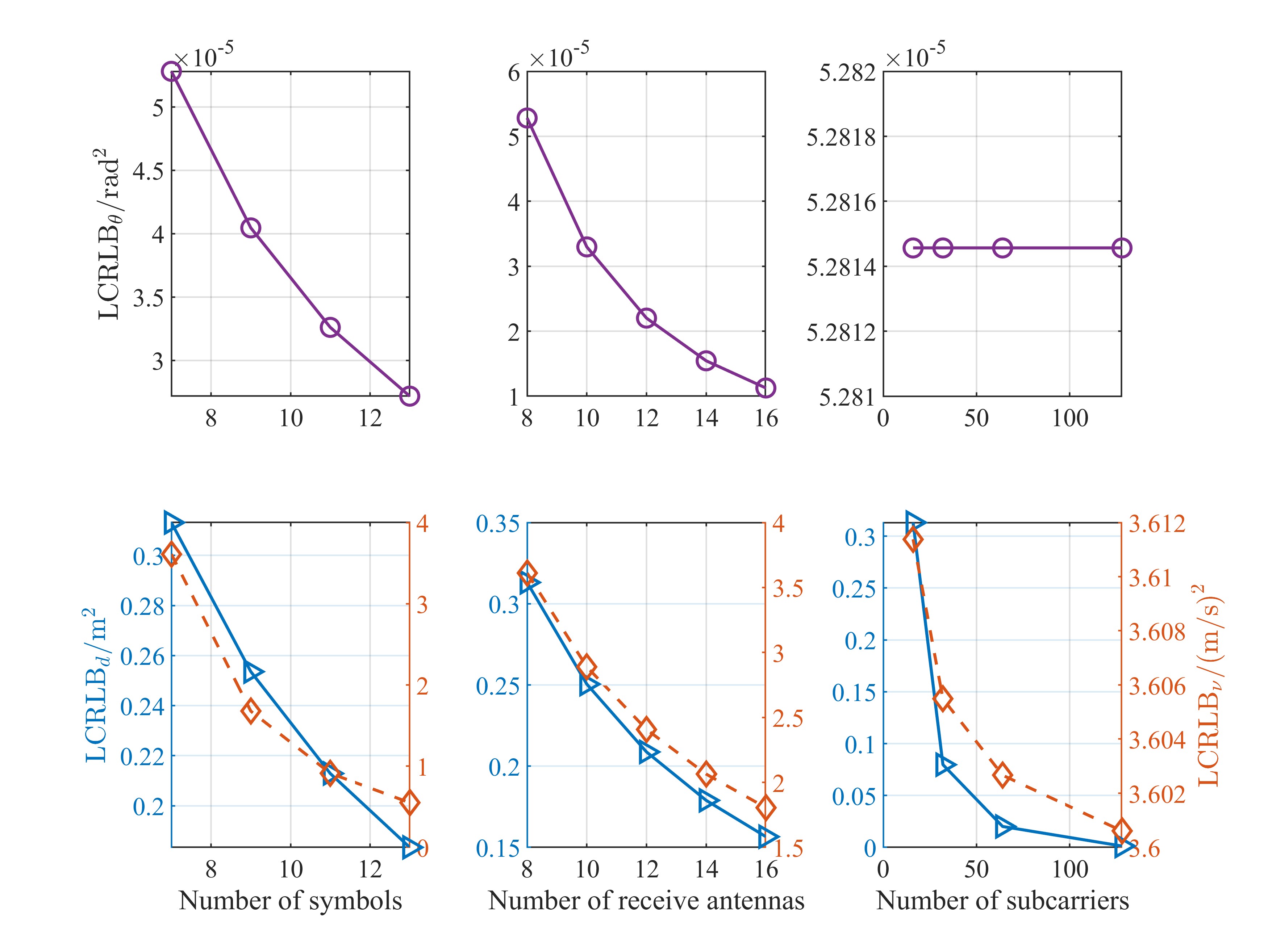}
\caption{\underline{\textbf{This figure is newly added}} The impact of system parameters on sensing LCRLBs under the $\rm{SNR=-5dB}$. 
We set the angle, distance, and velocity of vehicle as $[12^{\circ},410 \rm{m},18 \rm{m/s}]$, the numbers of symbols, receiving antennas, and subcarriers are 7, 8, and 16, respectively. 
When optimizing one variable, the other parameters remain unchanged.}
\label{fig2}
\end{figure}

\endgroup

\begingroup
\color{blue}

\subsection{Kinematics Model}
In dynamic scenarios, the study of kinematic characteristics is beneficial to the prediction of vehicle position.
The velocity model is constructed based on the magnitude of the corresponding velocity vectors, which is given by
\begin{equation}
\label{deqn_ex14a}
\nu_{k,m}=\nu_{k,m-1}+\Delta\nu_{k,m-1}, \forall k,m,
\end{equation}
where $\nu_{k,m}$ is the average velocity magnitude of the $k$-th vehicle at the $m$-th time slot and $\Delta\nu_{k,m-1}\sim \mathcal{U}(\Delta\nu_{min},\Delta\nu_{max})$ is the velocity increment.
According to the kinematic model described in \cite{ref2}, the geometric relations of movement behavior can be expressed as
\begin{equation}
\label{deqn_ex15a}
\left\{\begin{matrix}
sin(\theta_{k,m}-\theta_{k,m-1})d_{k,m}=\nu_{k,m-1}\Delta Tsin\theta_{k,m-1}, \\
d_{k,m}^{2}=d_{k,m-1}^{2}+(\Delta d)^{2}-2d_{k,m-1}\Delta dcos\theta_{k,m-1},
\end{matrix}\right.
\end{equation}
where $\Delta d=\nu_{k,m-1}\Delta T$ denotes the distance traveled by the $k$-th vehicle in the $(m-1)$-th time slot.
Based on this, the kinematic equation of vehicles can be modeled as
\begin{equation}
\left\{\begin{matrix}
\label{deqn_ex16a}
\begin{aligned}
\theta_{k,m}&=\theta_{k,m-1}+d_{k,m-1}^{-1}\nu_{k,m-1}\Delta Tsin\theta_{k,m-1}+\omega _{\theta},\\
d_{k,m}&=d_{k,m-1}-\nu_{k,m-1}\Delta Tcos\theta_{k,m-1}+\omega _{d},\\
\nu_{k,m}&=\nu_{k,m-1}+\Delta \nu_{k,m-1}+\omega _{\nu},
\end{aligned}
\end{matrix}\right.
\end{equation}
where $\omega _{\theta}$, $\omega _{d}$ and $\omega _{\nu}$ denote mutually independent zero-mean Gaussian noise with variances $\sigma_{\theta}^{2}$, $\sigma_{d}^{2}$ and $\sigma_{\nu}^{2}$, respectively. 
The corresponding noises are generated by a series of approximations of the motion trajectory and are unrelated to thermal noise.
The parameters of the $m$-th time slot can be calculated from the estimated parameters of the $(m-1)$-th time slot.
\endgroup

\begingroup
\color{blue}
\subsection{Vehicle Tracking via Extended Kalman Filtering}
For state variables $\bm{\zeta}$, the general representation of the kinematic evolution model and measurement model for the $m$-th time slot can be expressed as
\begin{equation}
\label{deqn_ex17a}
 \left\{\begin{matrix}
 \begin{aligned}
 \bm{\zeta}_{m}&=\boldsymbol{g}(\bm{\zeta}_{m-1})+\bm{\omega }_{m},\\
 \tilde{\mathbf{y}}_{m}&=\boldsymbol{h}(\bm{\zeta}_{m})+\tilde{\mathbf{z}}_{m}=\tilde{\mathbf{x}}_{m}+\tilde{\mathbf{z}}_{m},
 \end{aligned}
\end{matrix}\right.
\end{equation}
where $\boldsymbol{g}(\cdot)$ is the state function vectors determined by $(\ref{deqn_ex16a})$ and $\boldsymbol{h}(\cdot)$ is the measurement function vectors determined by $(\ref{deqn_ex9a})$. 
$\bm{\omega}=[\bm{\omega}_{1}^{T},\bm{\omega}_{2}^{T},\cdots,\bm{\omega}_{K}^{T}]^{T}$ and $\bm{\omega}_{k}=[\omega_{\theta},\omega_{d},\omega_{\nu}]^{T}$ is the zero-mean Gaussian noise vector, with covariance matrices being $\mathbf{\Sigma}_{\bm{\zeta}_{k}}=\mathrm{diag}(\sigma_{\theta}^{2},\sigma_{d}^{2},\sigma_{\nu}^{2})$.
Obviously, the kinematic evolution model and measurement model are nonlinear, the EKF approach can be used for beam tracking and state prediction, the overall designs are summarized as follows:

\textit{1) State Prediction:}
\begin{equation}
\label{deqn_ex18a}
\bm{\hat{\zeta}}_{m|m-1}=\boldsymbol{g}(\bm{\hat{\zeta}}_{m-1}).
\end{equation}

\textit{2) Linearization:}
\begin{equation}
\label{deqn_ex19a}
\mathbf{G}_{m-1}=\frac{\partial \boldsymbol{g}}{\partial \bm{\zeta}}|_{\bm{\zeta}=\bm{\hat{\zeta}}_{m-1}}, \mathbf{H}_{m}=\frac{\partial \boldsymbol{h}}{\partial \bm{\zeta}}|_{\bm{\zeta}=\bm{\hat{\zeta}}_{m|m-1}}.
\end{equation}

\textit{3) MSE Matrix Prediction:}
\begin{equation}
\label{deqn_ex20a}
\mathbf{\Theta}_{m|m-1}=\mathbf{G}_{m-1}\mathbf{\Theta}_{m-1}\mathbf{G}_{m-1}^{H}+ \mathbf{\Sigma}_{\bm{\zeta}}.
\end{equation}

\textit{4) Kalman Gain Calculation:}
\begin{equation}
\label{deqn_ex21a}
\mathbf{K}_{m}=\mathbf{\Theta}_{m|m-1}\mathbf{H}_{m}^{H}(\mathbf{\Sigma}+\mathbf{H}_{m}\mathbf{\Theta}_{m|m-1}\mathbf{H}_{m}^{H})^{-1}.
\end{equation}

\textit{5) State Tracking:}
\begin{equation}
\label{deqn_ex22a}
\bm{\hat{\zeta}}_{m}=\bm{\hat{\zeta}}_{m|m-1}+\bm{\mathrm{K}}_{m}(\mathbf{\tilde{y}}_{m}-\boldsymbol{h}(\bm{\hat{\zeta}}_{m|m-1})).
\end{equation}

\textit{6) MSE Matrix Update:}
\begin{equation}
\label{deqn_ex23a}
\mathbf{\Theta}_{m}=(\mathbf{I}-\mathbf{K}_{m}\mathbf{H}_{m})\mathbf{\Theta}_{m|m-1}.
\end{equation}
According to (\ref{deqn_ex16a}) and (\ref{deqn_ex9a}), the Jacobian matrix of $\boldsymbol{g}(\bm{\zeta})$ and $\boldsymbol{h}(\bm{\zeta})$ can be expressed as
\begin{equation}
\label{deqn_ex24a}
\scalebox{0.87}{$
\begin{aligned}
\frac{\partial \boldsymbol{g}(\bm{\zeta})}{\partial \bm{\zeta} }=\mathrm{diag}\left \{\begin{bmatrix}
  1+\frac{\nu_{k}\Delta Tcos\theta_{k}}{d_{k}} & -\frac{\nu_{k}\Delta Tsin\theta_{k}}{d_{k}^{2}}  & \frac{\Delta Tsin\theta_{k}}{d_{k}} \\
 \nu_{k}\Delta Tsin\theta_{k}& 1 & -\Delta Tcos\theta_{k} \\
 0& 0&  1 
\end{bmatrix} \right \}_{k=1}^{K}.
\end{aligned}$}
\end{equation}
and $\frac{\partial \boldsymbol{h}(\bm{\zeta})}{\partial \bm{\zeta} }=[\frac{\partial \mathbf{\tilde{x}}}{\partial \mathbf{u}_{1} }Q_{11}^{T},\cdots,\frac{\partial \mathbf{\tilde{x}}}{\partial \mathbf{u}_{K} }Q_{KK}^{T}]$ with $\frac{\partial \mathbf{\tilde{x}}}{\partial \mathbf{u}_{k} }Q_{kk}^{T}=\boldsymbol{\varpi}^{k}Q_{kk}^{T}$.
The initial vehicle motion parameters can be derived from the received sensing echoes, for example, by applying tensor decomposition \cite{ref29} to obtain coarse estimates suitable for initialization.
\endgroup

\begingroup
\color{blue}
\subsection{PCRLB for Parameter Estimation}
For parameter estimation in dynamic scenarios, the calculation of CRLB not only depends on the measurement model but is also affected by the state model. Naturally, the PCRLB is introduced. Following the definition in \cite{ref4}, the prior PDF
for $\bm{\zeta}$ at the $m$-th time slot can be expressed as
\begin{equation}
\scalebox{0.98}{$
\label{deqn_ex25a}
p(\bm{\zeta}_{m})=\frac{1}{\pi^{3}\mathrm{det}(\mathbf{\Theta}_{m|m-1})}e^{-(\bm{\zeta}_{m}-\boldsymbol{g}^{H}(\bm{\hat{\zeta}}_{m-1}))\mathbf{\Theta }_{m|m-1}^{-1}(\bm{\zeta}_{m}-\boldsymbol{g}(\bm{\hat{\zeta}}_{m-1}))}.$}
\end{equation}
Therefore, the FIM with respect to $p(\bm{\zeta}_{m})$ is given by
\begin{equation}
\label{deqn_ex26a}
\mathbf{J}_{m}^{P}=\mathbf{\Theta}_{m|m-1}^{-1} =(\mathbf{G}_{m-1}\mathbf{\Theta}_{m-1}\mathbf{G}_{m-1}^{H}+ \mathbf{\Sigma}_{\bm{\zeta}})^{-1}.
\end{equation}
According to \cite{ref4}, the posterior FIM (PFIM) is the sum of observed FIM and the prior FIM, which can be written as
\begin{equation}
\label{deqn_ex27a}
\mathbf{J}_{m}=\mathbf{J}_{m}^{O}+\mathbf{J}_{m}^{P},
\end{equation}
where $\mathbf{J}_{m}^{O}=\mathbf{J}(\bm{\zeta}_{m})$. Hence, the PCRLB is given by
\begin{equation}
\label{deqn_ex28a}
\mathrm{PCRLB}=\mathbf{J}_{m}^{-1}=(\mathbf{J}_{m}^{O}+\mathbf{J}_{m}^{P})^{-1}.
\end{equation}
It is worth noting that $\mathbf{J}_{m}^{O}$ is an Hermitian matrix, $\mathbf{J}_{m}^{P}$ is a real symmetric matrix, then $\mathbf{J}_{m}$ is an Hermitian matrix. 
Similarly, according to the property in \cite{ref28}, we can deduce that
\begin{equation}
\label{deqn_ex29a}
    [\mathbf{J}_{m}^{-1}]_{i+3(k-1),i+3(k-1)}\ge [(\mathbf{J}_{m,kk})^{-1}]_{ii},i=1,2,3.
\end{equation}

In addition, $\mathbf{J}_{m}^{P}$ is a block diagonal matrix, which follows $\mathbf{J}_{m}^{P}=\mathrm{diag}(\bm{\mathcal{J}}_{kk,m}^{P})_{k=1}^{K}$.
Thus, $\mathbf{J}_{m,kk}$ can be expressed as
\begin{equation}
\label{deqn_ex30a}
\scalebox{0.9}{$
\begin{aligned}
    &\mathbf{J}_{m,kk}=\bm{\mathcal{Q}}_{kk,m}+ \bm{\mathcal{J}}_{kk,m}^{P}=\\&\left [ \begin{matrix}
  \bm{\mathcal{Q}}_{kk,m}^{11}+\bm{\mathcal{J}}_{kk,m}^{11}& \bm{\mathcal{Q}}_{kk,m}^{12}+\bm{\mathcal{J}}_{kk,m}^{12} & \bm{\mathcal{Q}}_{kk,m}^{13}+\bm{\mathcal{J}}_{kk,m}^{13}\\
 \bm{\mathcal{Q}}_{kk,m}^{21}+\bm{\mathcal{J}}_{kk,m}^{21} & \bm{\mathcal{Q}}_{kk,m}^{22}+\bm{\mathcal{J}}_{kk,m}^{22} & \bm{\mathcal{Q}}_{kk,m}^{23}+\bm{\mathcal{J}}_{kk,m}^{23}\\
\bm{\mathcal{Q}}_{kk,m}^{31}+\bm{\mathcal{J}}_{kk,m}^{31}  &\bm{\mathcal{Q}}_{kk,m}^{32}+\bm{\mathcal{J}}_{kk,m}^{32}  &\bm{\mathcal{Q}}_{kk,m}^{33}+\bm{\mathcal{J}}_{kk,m}^{33}
\end{matrix} \right ], 
    \end{aligned}$}
\end{equation}
where $\bm{\mathcal{J}}_{kk,m}^{ij}$ represents the $(i,j)$-th element of $\bm{\mathcal{J}}_{kk,m}^{P}$.
Due to the complexity of performance analysis caused by the inverse of matrix $\mathbf{J}_{m,kk}$, we plan to ignore the elements in the non-principal diagonal blocks of $\mathbf{J}_{m,kk}$ to obtain the LPCRLB. 
The LPCRLB matrices can be expressed as
\begin{equation}
\label{deqn_ex31a}
\scalebox{0.76}{$
\begin{aligned}
&\mathrm{LPCRLB}_{k,m}^{\theta}=\\&\big( sin^{2}\theta_{k,m} \mathbf{p}^{T}\mathbf{g}_{kk,m}^{11} +\frac{4\nu_{k,m}sin^{2}\theta_{k,m}}{\lambda}(\mathbf{p}^{T}\mathbf{g}_{kk,m}^{31} +\frac{\nu_{k,m}}{\lambda}\mathbf{p}^{T}\mathbf{g}_{kk,m}^{33})+\bm{\mathcal{J}}_{kk,m}^{11}\big)^{-1},\\
&\mathrm{LPCRLB}_{k,m}^{d\nu}=\begin{bmatrix}\frac{4}{c^{2}}\mathbf{p}^{T}\mathbf{g}_{kk,m}^{22}+\bm{\mathcal{J}}_{kk,m}^{22} & \frac{4cos\theta_{k.m}}{c\lambda}\mathbf{p}^{T}\mathbf{g}_{kk,m}^{23} +\bm{\mathcal{J}}_{kk,m}^{23}\\\frac{4cos\theta_{k,m}}{c\lambda}\mathbf{p}^{T}\mathbf{g}_{kk,m}^{32} +\bm{\mathcal{J}}_{kk,m}^{32}  &\frac{4cos^{2}\theta_{k,m}}{\lambda^{2}}\mathbf{p}^{T}\mathbf{g}_{kk,m}^{33}+\bm{\mathcal{J}}_{kk,m}^{33}
    \end{bmatrix}^{-1}.
\end{aligned}$}
\end{equation}
Further, the LPCRLBs of distance and velocity can be written as  
\begin{equation}
\label{deqn_ex32a}
\scalebox{0.7}{$
\begin{aligned}
    \mathrm{LPCRLB}_{k,m}^{(-1)d}&=(\frac{4}{c^{2}}\mathbf{p}^{T}\mathbf{g}_{kk,m}^{22}+\bm{\mathcal{J}}_{kk,m}^{22})-\frac{(\frac{4cos\theta_{k.m}}{c\lambda}\mathbf{p}^{T}\mathbf{g}_{kk,m}^{23} +\bm{\mathcal{J}}_{kk,m}^{23})^{2}}{\frac{4cos^{2}\theta_{k,m}}{\lambda^{2}}\mathbf{p}^{T}\mathbf{g}_{kk,m}^{33}+\bm{\mathcal{J}}_{kk,m}^{33}},\\
    \mathrm{LPCRLB}_{k,m}^{(-1)\nu}&=(\frac{4cos^{2}\theta_{k,m}}{\lambda^{2}}\mathbf{p}^{T}\mathbf{g}_{kk,m}^{33}+\bm{\mathcal{J}}_{kk,m}^{33})-\frac{(\frac{4cos\theta_{k.m}}{c\lambda}\mathbf{p}^{T}\mathbf{g}_{kk,m}^{23} +\bm{\mathcal{J}}_{kk,m}^{23})^{2}}{\frac{4}{c^{2}}\mathbf{p}^{T}\mathbf{g}_{kk,m}^{22}+\bm{\mathcal{J}}_{kk,m}^{22}}.
\end{aligned}$}
\end{equation}
Observing $(\ref{deqn_ex61a})$ and $(\ref{deqn_ex62a})$, it can be seen that the $\mathrm{LPCRLB}$ of DoD is related to the positions of the transmit antenna and receive antenna, while the $\mathrm{LPCRLBs}$ of distance and velocity are only related to the position of the transmit antenna.
In addition, the sensing estimation accuracy of all motion parameters is related to the transmit power.
\endgroup

{\color{blue} \subsection{Communication Model}
For the $m$-th time slot, the} communication signal received by the $k$-th vehicle on the $n$-th subcarrier in the $q$-th block is \cite{ref18}
\begin{equation}
\label{deqn_ex33a}
{\color{blue}
y_{k,n,m}^{q}=\mathbf{h}_{k,n,m}^{\left(q\right)H}\mathbf{w}_{n}^{m}c_{n,m}^{q}+\eta_{k,n,m}^{q},} n\in\mathcal{N}_{k},
\end{equation}
where {\color{blue}$\mathbf{h}_{k,n,m}^{(q)}=\sqrt{\alpha_{k,m}}e^{-j\varphi_{k,m}}\mathbf{a}(\mathbf{p}_{tx}^{m},\theta_{k,m})$ with $\varphi_{k,m}$ is the phase shift caused by the delay and the Doppler frequency of the $k$-th vehicle. 
Since the vehicle's own velocity $\nu_{k,m}$ is known and the arrival angle $\theta_{k,m}$ can be estimated based on the angle prediction received from the RSU, the Doppler shift can easily be compensated at the vehicle’s receiver.
$\alpha_{k,m}=\alpha_{0}\left(\frac{d_{k,n,m}}{d_{0}}\right)^{-\xi }$} is the path loss coefficient, where $\alpha_{0}$ is the path loss at reference distance $d_{0}$, {\color{blue}$d_{k,n,m}$} represents the distance between the $k$-th vehicle and the RSU on the $n$-th subcarrier, and $\xi $ is the associated path loss exponent. 
${\color{blue}\eta_{k,n,m}^{q}\sim \mathcal{CN}\left(0,\eta_{0}/T_{e}\right) }$  denotes the noise in the $k$-th vehicle, with {\color{blue}$\eta_{0}/T_{e}$} being the variance of noise.
Thus, the signal-to-noise ratio (SNR) received in the $k$-th vehicle can be expressed as
\begin{equation}
\label{deqn_ex34a}
\scalebox{0.97}{$
{\color{blue}\mathrm{S}_{k,n,m}=\left|\mathbf{h}_{k,n,m}^{\left(q\right)H}\mathbf{w}_{n}^{m}\right|^2T_{e}/\eta_{0}}, n\in\mathcal{N}_{k}.$}
\end{equation}

It is noted that {\color{blue}$\mathrm{S}_{k,n,m}$} depends on APV {\color{blue}$\mathbf{p}_{tx}^{m}$} and {\color{blue}$\mathbf{w}_{n}^{m}$}. 
Therefore, the transmission rate between the transmitter and the $k$-th vehicle is given by
\textcolor{blue}{
\begin{equation}
\label{deqn_ex35a}
\scalebox{0.95}{$
\mathrm{R}_{k.m}=\sum\limits_{n\in\mathcal{N}_{k}}\mathrm{log}\left(1+\alpha_{k,m}\left|\mathbf{a}^{H}(\mathbf{p}_{tx}^{m},\theta_{k,m})\mathbf{w}_{n}^{m}\right|^2p_{n,m}T_{e}/\eta_{0}\right),$}
\end{equation}
where $p_{n,m}=\mathbb{E}\{\left|c_{n.m}^{q}\right|^{2}\}$ is the transmit power on the $n$-th subcarrier. It is noted that {\color{blue}$\mathrm{R}_{k,m}$} depends on the transmit APV and power.}

\section{Joint Beamforming and Antenna Position Optimization Modeling}
In this section, {\color{blue}we first analyze the challenges faced by dynamic scenarios and then propose two optimization frameworks to address them. }

\begingroup
\color{blue}
\subsection{Problem Description}
In high-mobility vehicular scenarios, real-time beam alignment is typically difficult to maintain, and predictive beamforming is therefore employed to continuously track the vehicle. However, the rapid motion of the vehicle results in fast time-varying channel conditions, under which the instantaneous channel state may fail to provide desirable communication or sensing performance. Consequently, proactively creating and maintaining favorable channel conditions becomes essential, as the instantaneous channel quality directly determines the achievable communication rate and sensing accuracy within each time slot.
To address these challenges, we develop two complementary optimization frameworks from both theoretical and practical perspectives, enabling a systematic analysis of how channel variations affect joint communication and sensing performance.

In the theoretical setting without QoS requirements, we formulate a weighted-sum maximization problem for communication sum-rate and sensing LPCRLB. By adjusting the weighting factor $\rho$, the inherent trade-off between communication sum-rate and sensing accuracy can be characterized, which further reveals the feasible performance region of the system.
In practical systems, however, sensing must satisfy a minimum performance requirement. Therefore, when a sensing QoS constraint is imposed, we further develop a communication sum-rate maximization problem subject to a prescribed CRLB threshold, aiming to enhance the communication performance while guaranteeing the required sensing capability.

It is worth noting that the feasible sensing performance region obtained from the weighted-sum maximization problem is used to determine the LPCRLB threshold in the sensing QoS-constrained problem. In this way, the two problems form a complementary pair: the first explores the achievable communication–sensing performance boundary, while the second enables practical operation under required sensing QoS.
\endgroup

\subsection{Problem Formulation}
To suppress the coupling effect between the adjacent antenna elements in the flexible array, 
each antenna pair needs to satisfy a minimum distance constraint, namely,
{\color{blue}
\begin{equation}
\label{deqn_ex36a}
\scalebox{0.9}{$
\begin{aligned}
&\left\| p_{tx,l}- p_{tx,l^{'}} \right \|_{2}\ge D_{sp}, \forall l\ne l^{'},
{\color{blue}\left\| p_{rx,ll} - p_{rx,ll^{'}} \right \|_{2}}\ge D_{sp},\forall ll\ne ll^{'},
\end{aligned}$}
\end{equation}}
where {\color{blue}$D_{sp}$} represent the minimum distance between adjacent antenna elements, respectively. 
In addition, the antenna spacing $D_{tr}$ is set to ensure physical isolation between the transceiver, satisfying $D_{min}^{rx}-D_{max}^{tx}=D_{tr}$.

{\color{blue} 
For the $m$-th time slot, performance optimization is often expected based on the current channel conditions and estimated parameters.
However, in dynamic scenarios, real-time data processing and optimization are not feasible.
Inspired by the concept of predictive beamforming, we utilize the motion parameters $\bm{\hat{\zeta}}_{m|m-1}$ predicted by the EKF at the $m$-th time slot as the reference, where the expected value is taken as the estimated channel for computing the communication sum-rate, while one of the predicted results is used to construct the sensing channel for calculating the sensing LPCRLBs.}

Based on the derived {\color{blue}LPCRLBs} of motion parameters estimation, defining the matrix ${\color{blue}\bm{\mathrm{L}}_{k,m}=\mathrm{diag}([      
\mathrm{LPCRLB}_{{\hat{\theta}}_{k,m|m-1}},}\\{\color{blue}\mathrm{LPCRLB}_{{\hat{d}}_{k,m|m-1}},\mathrm{LPCRLB}_{{\hat{\nu}}_{k,m|m-1}}
])}$, the {\color{blue} weighted-sum maximization} optimization problem can be formulated as
\begin{subequations}
\begin{align}
\mathrm{(P1)}\quad&\mScale{\max_{{\color{blue}\mathbf{p}_{tx}^{m},\mathbf{p}_{rx}^{m},\mathbf{p}^{m},\mathbf{w}^{m}}} \rho {\color{blue}\sum_{k=1}^{K}\mathrm{R}_{k,m}}+\left ( 1-\rho \right )\sum_{k=1}^{K}\sum_{i=1}^{{\color{blue}3}} \frac{1}{{\color{blue}\bm{\mathrm{L}}_{k,m}^{ii}}},}\nonumber \\
\mathrm{\mathrm{s.t.}}\quad&\mScale{{\color{blue}\left\| p_{tx,l}^{m} - p_{tx,l^{'}}^{m} \right \|_{2}\ge D_{sp}},\quad\forall l\ne l^{'},}\label{deqn_ex37A}\\
\quad&\mScale{{\color{blue}\left\| p_{rx,ll}^{m} - p_{rx,ll^{'}}^{m} \right \|_{2}\ge D_{sp}},\quad\forall ll\ne ll^{'},}\label{deqn_ex37B}\\
\quad &\mScale{D_{min}^{tx}\le p_{tx,1}^{{\color{blue}m}}<\cdots < p_{tx,M_{tx}}^{{\color{blue}m}}\le D_{min}^{rx}-D_{tr},}\label{deqn_ex37C}\\
\quad\quad\quad &\mScale{D_{max}^{tx}+D_{tr} \le p_{rx,1}^{{\color{blue}m}}<\cdots < p_{rx,M_{rx}}^{{\color{blue}m}}\le D_{max}^{rx},}\label{deqn_ex37D}\\
\quad &\mScale{\sum\nolimits_{n=1}^{N} {\color{blue} p_{n}^{m}} \le  P_{T} ,\quad n\in\mathcal{N}_{k}.}\label{deqn_ex37E}\\
\quad &\mScale{ {\color{blue} \left | \mathbf{w}_{n}^{m} \right | } =1,\quad n\in\mathcal{N}_{k},}\label{deqn_ex37F}
\end{align}
\end{subequations}
The constraints (\ref{deqn_ex37C}) and (\ref{deqn_ex37D}) restrict the MA to move within the feasible region {\color{blue}$\left[D_{min}^{tx},D_{max}^{tx}\right]$ and $\left[D_{min}^{rx},D_{max}^{rx}\right]$}.
Constraint {\color{blue}(\ref{deqn_ex37E})} is the limited transmit power constraint.  

\begingroup
\color{blue}
As discussed in Section II-E, the position of the receive antennas mainly affects the DoD sensing accuracy, and optimizing them would further tighten the angular constraint, potentially degrading the communication performance. 
Therefore, for the problem of maximizing the sum-rate under the sensing LPCRLB constraints, we focus exclusively on optimizing the position of the transmit antennas.
The optimization problem can be written as
\begin{subequations}
\begin{align}
\mathrm{(P2)}\quad&\mScale{{\color{blue}\min_{\mathbf{p}_{tx}^{m},\mathbf{p}^{m},\mathbf{w}^{m}}\quad\quad-\sum_{k=1}^{K}\mathrm{R}_{k,m},}}\nonumber \\
\mathrm{\mathrm{s.t.}}\quad&\mScale{{\color{blue}\left\| p_{tx,l}^{m} - p_{tx,l^{'}}^{m}  \right \|_{2}}\ge D_{sp},\quad\forall l\ne l^{'},}\label{deqn_ex38A}\\
\quad &\mScale{D_{min}^{tx}\le p_{tx,1}^{{\color{blue}m}}<\cdots < p_{tx,M_{tx}}^{{\color{blue}m}}\le D_{min}^{rx}-D_{tr},}\label{deqn_ex38B}\\
\quad &\mScale{{\color{blue}\mathrm{LPCRLB}_{{\hat{\theta}}_{k,m|m-1}}\le \varsigma_{k}^{\theta},}}\label{deqn_ex38C}\\
\quad &\mScale{{\color{blue}\mathrm{LPCRLB}_{{\hat{d}}_{k,m|m-1}}\le \varsigma_{k}^{d},}}\label{deqn_ex38D}\\
\quad &\mScale{{\color{blue}\mathrm{LPCRLB}_{{\hat{\nu}}_{k,m|m-1}}\le \varsigma_{k}^{\nu},}}\label{deqn_ex38E}\\
\quad &\mScale{{\color{blue}\sum\nolimits_{n=1}^{N} p_{n}^{m} \le  P_{T},\quad n\in\mathcal{N}_{k}.}}\label{deqn_ex38F}\\
\quad &\mScale{{\color{blue}\left | \mathbf{w}_{n}^{m} \right | =1,\quad n\in\mathcal{N}_{k},}}\label{deqn_ex38G}
\end{align}
\end{subequations}
where $\varsigma_{k}^{\theta},\varsigma_{k}^{d},\varsigma_{k}^{\nu}$ 
denote the LPCRLB constraints corresponding to the estimations of the DoD, distance, and velocity, respectively, whose values are assumed to be known.
To highlight the core structure of the proposed algorithms, we omit the time-slot subscripts to avoid unnecessary notational clutter in the following sections.
\endgroup

\begin{figure}[!t]
\centering
\includegraphics[width=3.5in]{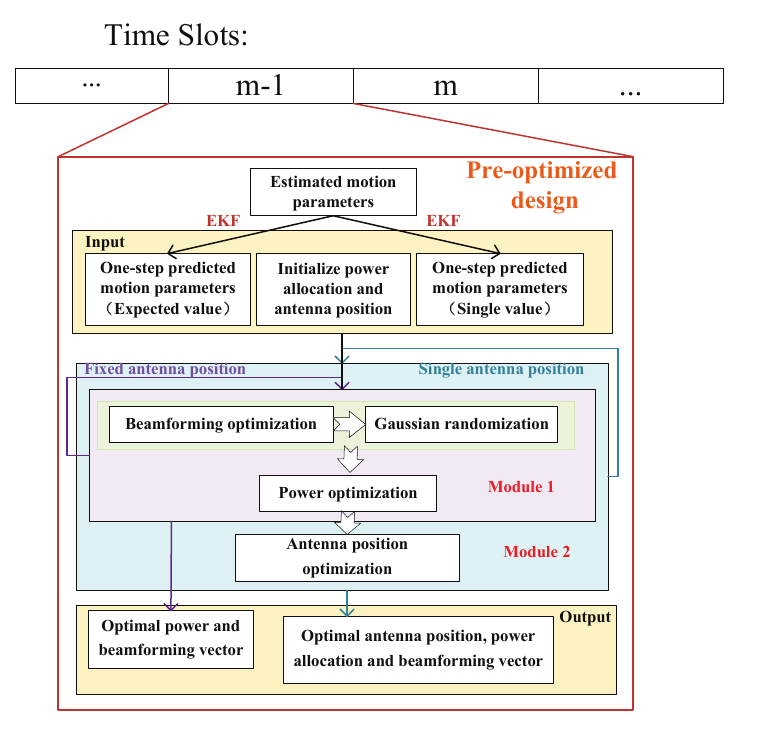}
\caption{\underline{\textbf{This figure is newly added}} The PRE-SC-PGA algorithm framework for weighted-sum maximization.}
\label{fig3}
\end{figure}

\begingroup
\color{blue}
\section{PRE-SC-PGA Algorithm Design For Weighted-Sum Maximization Problem}
This section focuses on algorithm design for the weighted maximization problem in scenarios without quality of service constraints.
The optimization variables are highly coupled within the objective function, making the problem difficult to solve directly. 
Therefore, we design an alternating optimization (AO) algorithm called PRE-SC-PGA, which iteratively optimizes each parameter to obtain a suboptimal solution to the problem (P1). 

\subsection{PRE-SC-PGA Algorithm Framework Design}
To anticipate the impact of channel variations on communication and sensing performance in the $m$-th time slot, we use the estimated parameters from the $(m-1)$-th time slot as a basis to predict the vehicle's motion state in the next time slot using EKF, thus providing the communication and sensing estimation channel for the $m$-th time slot.
As illustrated in Fig.\ref{fig3}, based on the designed channel conditions and given the initialized antenna position vectors and transmit power, the proposed PRE-SC-PGA algorithm (Module 2) is applied to solve problem (P1), achieving joint optimization of communication and sensing performance.
The sensing coverage obtained through this process can provide prior information for the sensing LPCRLBs of the subsequent communication sum-rate maximization problem.
It is worth noting that when the optimization variable is the power vector, the corresponding objective function and constrain is convex and can be optimally solved using the CVX solver \cite{ref30}. 
Therefore, this section primarily focuses on developing optimization algorithms for the beamforming vector $\mathbf{w}$ and the APVs $\mathbf{p}_{tx}$ and $\mathbf{p}_{rx}$.
\endgroup

\subsection{Preprocessing}
Taking into account {\color{blue}the different contributions of the magnitudes of communication sum-rate and sensing LPCRLBs}, we introduce quantization factors ${\color{blue} \left\{\aleph_{i}\right\}_{i=1}^{3}}$ to eliminate the ``biased'' effect. 
After preprocessing, {\color{blue}the sensing components of the objective function in} $\color{blue} \mathrm{(P1)}$ can be restated in the following forms
\begingroup
\color{blue}
\begin{equation}
\begin{aligned}
\label{deqn_ex39a}
 \sum_{k=1}^{K}\sum_{i=1}^{3} \frac{1}{\bm{\mathrm{L}}_{k}^{ii}} = \sum_{k=1}^{K}\sum_{i=1}^{3} \aleph_{i}\frac{1}{\bm{\mathrm{L}}_{k}^{ii}}.
\end{aligned}
\end{equation}
\endgroup
\subsection{Beamforming Design}
{\color{blue}
By initializing the position vectors $\mathbf{p}_{tx}$ and $\mathbf{p}_{rx}$, and given the transmit power vector obtained from the uniform allocation, the problem $\mathrm{(P1)}$ can be reformulated as}
\begin{equation}
{\color{blue}
\begin{aligned}
\label{deqn_ex40a}
\mathrm{(P1.1)}\quad\max_{\mathbf{w}}&\quad  \sum_{k=1}^{K}\rho\mathrm{R}_{k}+\left ( 1-\rho \right )\sum_{k=1}^{K}\sum_{i=1}^{3} \aleph_{i}\frac{1}{\bm{\mathrm{L}}_{k}^{ii}},\\
\mathrm{s.t.}&\quad\quad\quad\quad\quad\quad\quad \rm{(\ref{deqn_ex37F})}.
\end{aligned}}
\end{equation}
To solve this problem, we first deal with constraints based on the idea of semidefinite relaxation, 
and then use successive convex approximation (SCA) to approximate the relaxed problem into a convex problem. 
For the problem of constraint handling, 
we define {\color{blue}$\mathbf{W}_{n}=\mathbf{w}_{n}\mathbf{w}_{n}^{H}$, 
where $\mathbf{W}_{n}$} is positive semidefinite, and $\mathrm{rank}({\color{blue}\mathbf{W}_{n}})=1$. 
Based on the definition, we can derive that ${\color{blue}\mathbf{w}_{n}^{H}}\bm{\bm{\mathcal{K}}}_{k}{\color{blue}\mathbf{w}_{n}}=\mathrm{Tr}(\bm{\mathcal{K}}_{k}{\color{blue}\mathbf{W}_{n}})$, ${\color{blue}\mathbf{w}_{n}^{H}}\bm{\bm{\mathcal{G}}}_{k}{\color{blue}\mathbf{w}_{n}}=\mathrm{Tr}(\bm{\mathcal{G}}_{k}{\color{blue}\mathbf{W}_{n}})$, ${\color{blue}\mathbf{w}_{n}^{H}}\mathbf{\Lambda}_{tx} \bm{\mathcal{G}}_{k}{\color{blue}\mathbf{w}_{n}}=\mathrm{Tr}\left(\mathbf{\Lambda}_{tx}\bm{\mathcal{G}}_{k}{\color{blue}\mathbf{W}_{n}}\right)$.
Due to the non-negativity of {\color{blue}$\bm{\mathrm{L}}_{k}^{ii}$}  , 
by introducing the auxiliary variable sets $\boldsymbol{\kappa}=\left\{\kappa_{1},\kappa_{2},\dots,\kappa_{K}\right\}$, $\boldsymbol{\epsilon}=\left\{\epsilon_{1},\epsilon_{2},\dots,\epsilon_{K}\right\}$, $\boldsymbol{\varepsilon}=\left\{\varepsilon_{1},\varepsilon_{2},\dots,\varepsilon_{K}\right\}$ and using SC, 
the optimization subproblem is equivalent to 
\begin{subequations}
\begin{align}
&\mScale{{\color{blue}\mathrm{(P1.2)}\quad \max_{\mathbf{W}, \boldsymbol{\kappa}, \boldsymbol{\epsilon},\boldsymbol{\varepsilon}}\,\,\,\rho  \mathcal{F}_{1}(\mathbf{W}) + \left(1- \rho \right)\mathcal{F}_{2}( \boldsymbol{\kappa}, \boldsymbol{\epsilon},\boldsymbol{\varepsilon})}}\nonumber\\
\mathrm{s.t.}\,\,&\mScale{{\color{blue}\mathbf{W}_{n}}\succeq 0,}\label{deqn_ex41A} \\
&\mScale{\mathrm{rank}({\color{blue}\mathbf{W}_{n}})=1,}\label{deqn_ex41B}\\
&\mScale{{\color{blue}\mathbf{W}_{n,n}=1,}}\label{deqn_ex41C}\\
&\mScale{{\color{blue}\bm{\mathcal{Q}}_{kk}^{11}+\bm{\mathcal{J}}_{kk}^{11}>=\kappa_{k},}}\label{deqn_ex41D}\\
&\mScale{{\color{blue}\begin{bmatrix}\frac{4}{c^{2}}\mathbf{p}^{T}\mathbf{g}_{kk}^{22}+\bm{\mathcal{J}}_{kk}^{22}-\epsilon_{k} & \frac{4cos\theta_{k}}{c\lambda}\mathbf{p}^{T}\mathbf{g}_{kk}^{23} +\bm{\mathcal{J}}_{kk}^{23}\\\frac{4cos\theta_{k}}{c\lambda}\mathbf{p}^{T}\mathbf{g}_{kk}^{32} +\bm{\mathcal{J}}_{kk}^{32}  &\frac{4cos^{2}\theta_{k}}{\lambda^{2}}\mathbf{p}^{T}\mathbf{g}_{kk}^{33}+\bm{\mathcal{J}}_{kk}^{33}
    \end{bmatrix} \succeq 0,}}\label{deqn_ex41E}\\
&\mScale{{\color{blue}\begin{bmatrix}\frac{4cos^{2}\theta_{k}}{\lambda^{2}}\mathbf{p}^{T}\mathbf{g}_{kk}^{33}+\bm{\mathcal{J}}_{kk}^{33}-\varepsilon_{k} & \frac{4cos\theta_{k}}{c\lambda}\mathbf{p}^{T}\mathbf{g}_{kk}^{23} +\bm{\mathcal{J}}_{kk}^{23}\\\frac{4cos\theta_{k}}{c\lambda}\mathbf{p}^{T}\mathbf{g}_{kk}^{32} +\bm{\mathcal{J}}_{kk}^{32}  & \frac{4}{c^{2}}\mathbf{p}^{T}\mathbf{g}_{kk}^{22}+\bm{\mathcal{J}}_{kk}^{22}
    \end{bmatrix} \succeq 0,}\label{deqn_ex41F}}
\end{align}
\end{subequations}
{\color{blue}where $\mathcal{F}_{1}(\mathbf{W})= \sum_{k=1}^{K}\mathrm{R}_{k}$ and $\mathcal{F}_{2}( \boldsymbol{\kappa}, \boldsymbol{\epsilon},\boldsymbol{\varepsilon})=\sum_{k=1}^{K}(\aleph_{1}\kappa_{k}+\aleph_{2}\epsilon_{k}+\aleph_{3}\varepsilon_{k})$. 
Noting that 
$\mathbf{w}_{n}$ in $\mathbf{g}_{kk}$ has been replaced by $\mathbf{W}_{n}$.}

To facilitate problem solving, we remove the rank-one constraint to make $\mathrm{(P1.2)}$ a convex optimization problem, and use CVX \cite{ref30} to obtain the optimal value {\color{blue}$\mathbf{W}_{n}^{\star}$} with the rank satisfying {\color{blue}$\mathrm{rank}(\mathbf{W}_{n}^{\star})>1$}. 
The convergence analysis of the beamforming algorithm is shown in \textbf{Appendix B}. 
Then, Gaussian randomization is used to construct an approximate rank-one solution {\color{blue}$\hat{\mathbf{W}}_{n}$} to the problem. 
To be specific, we first generate some random realizations $\mathbf{x}\sim\mathcal{CN}(0,{\color{blue}\hat{\mathbf{W}}_{n}} )$ and construct a set of candidate feasible solutions, 
then select {\color{blue}$\hat{\mathbf{w}}_{n}^{\ast}$} that maximizes the objective function.

\subsection{Transmit Antenna Design}
{\color{blue} With variables $\mathbf{p}_{rx}$, $\mathbf{p}$ and $\mathbf{w}$ fixed, the optimization variable $\mathbf{p}_{tx}$ only appears implicitly in the term $\bm{\mathcal{Q}}_{kk}$. 
To facilitate optimization of the variable, we first reformulate these terms in an explicit form of $\mathbf{p}_{tx}$.}
Define $\mathbf{f}_{k}=[\mathrm{f}_{1,k},\mathrm{f}_{2,k},\cdots,\mathrm{f}_{M_{tx},k}]^T$, ${\color{blue}\mathbf{s}_{k}=[\mathrm{s}_{1,k},\mathrm{s}_{2,k},\cdots,\mathrm{s}_{M_{tx},k}]^T}$ for $k=1,2,\cdots,K$, where $\mathrm{f}_{l,k}=cos(2\pi p_{tx,l}cos\theta_{k}/\lambda)$ and ${\color{blue}\mathrm{s}}_{l,k}=sin(2\pi p_{tx,l}cos\theta_{k}/\lambda)$. 
{\color{blue} In addition, we convert the optimized variable $\mathbf{\hat{w}}_{n}^{\ast}$ into its complex form, and have $\mathbf{\hat{w}}_{n}^{\ast}$} $=\hbar+ j\eth$.
{\color{blue}Define} $\mathbf{A}=\mathbf{f}_{k}\mathbf{f}_{k}^{T}+{\color{blue}\mathbf{s}_{k}\mathbf{s}_{k}^{T}}$, ${\color{blue}\mathbf{U}}=\mathbf{f}_{k}{\color{blue}\mathbf{s}}_{k}^{T}-{\color{blue}\mathbf{s}}_{k}\mathbf{f}_{k}^{T}$, ${\color{blue}\mathbf{T}}=\hbar\hbar^{T}+\eth\eth^{T}$ and ${\color{blue}\mathbf{D}}=\hbar\eth^{T}-\eth\hbar^{T}$, the terms {\color{blue} $\hat{\mathbf{w}}_{n}^{\ast(H)}\bm{\mathcal{G}}_{k}\hat{\mathbf{w}}_{n}^{\ast},\hat{\mathbf{w}}_{n}^{\ast(H)}\bm{\bm{\mathcal{K}}}_{k}\hat{\mathbf{w}}_{n}^{\ast}, \hat{\mathbf{w}}_{n}^{\ast(H)}\mathbf{\Lambda}_{tx}\bm{\mathcal{G}}_{k}\hat{\mathbf{w}}_{n}^{\ast}$ and $\hat{\mathbf{w}}_{n}^{\ast(H)}\bm{\mathcal{G}}_{k}\mathbf{\Lambda}_{tx}\hat{\mathbf{w}}_{n}^{\ast}$ can be rewritten as $\Xi(\mathbf{f}_{k},\mathbf{s}_{k})$, $\Xi_{1}(\mathbf{f}_{k},\mathbf{s}_{k},\mathbf{\Lambda}_{tx})$, $\Xi_{2}(\mathbf{f}_{k},\mathbf{s}_{k},\mathbf{\Lambda}_{tx})$ and $\Xi_{3}(\mathbf{f}_{k},\mathbf{s}_{k},\mathbf{\Lambda}_{tx})$, respectively, which follows as} 
\begin{equation}
\label{deqn_ex42a}
\scalebox{0.85}{$
\begin{aligned}
\Xi(\mathbf{f}_{k},{\color{blue}\mathbf{s}}_{k})&=\mathbf{f}_{k}^{T}{\color{blue}\mathbf{T}}\mathbf{f}_{k}+{\color{blue}\mathbf{s}}_{k}^{T}{\color{blue}\mathbf{T}}{\color{blue}\mathbf{s}}_{k}+2\mathbf{f}_{k}^{T}{\color{blue}\mathbf{D}}{\color{blue}\mathbf{s}}_{k},\\
\Xi_{1}(\mathbf{f}_{k},{\color{blue}\mathbf{s}}_{k},\mathbf{\Lambda}_{tx})&=\mathbf{f}_{k}^{T}\mathbf{\Lambda}_{tx}{\color{blue}\mathbf{T}}\mathbf{\Lambda}_{tx}\mathbf{f}_{k}+{\color{blue}\mathbf{s}}_{k}^{T}\mathbf{\Lambda}_{tx}{\color{blue}\mathbf{T}}\mathbf{\Lambda}_{tx}{\color{blue}\mathbf{s}}_{k}+2\mathbf{f}_{k}^{T}\mathbf{\Lambda}_{tx}{\color{blue}\mathbf{D}}\mathbf{\Lambda}_{tx}{\color{blue}\mathbf{s}}_{k},\\
\Xi_{2}(\mathbf{f}_{k},{\color{blue}\mathbf{s}}_{k},\mathbf{\Lambda}_{tx})
&=\hbar^{T}\mathbf{\Lambda}_{tx}(\mathbf{A}-j{\color{blue}\mathbf{U}})\hbar+\eth^{T}\mathbf{\Lambda}_{tx}(\mathbf{A}-j{\color{blue}\mathbf{U}})\eth\\
&+\hbar^{T}\mathbf{\Lambda}_{tx}({\color{blue}\mathbf{U}}+j\mathbf{A})\eth-\eth^{T}\mathbf{\Lambda}_{tx}({\color{blue}\mathbf{U}}+j\mathbf{A})\hbar,\\
\Xi_{3}(\mathbf{f}_{k},{\color{blue}\mathbf{s}}_{k},\mathbf{\Lambda}_{tx})&=\hbar^{T}(\mathbf{A}-j{\color{blue}\mathbf{U}})\mathbf{\Lambda}_{tx}\hbar+\eth^{T}(\mathbf{A}-j{\color{blue}\mathbf{U}})\mathbf{\Lambda}_{tx}\eth\\
&+\hbar^{T}({\color{blue}\mathbf{U}}+j\mathbf{A})\mathbf{\Lambda}_{tx}\eth-\eth^{T}({\color{blue}\mathbf{U}}+j\mathbf{A})\mathbf{\Lambda}_{tx}\hbar,
\end{aligned}$}
\end{equation}
Therefore, problem $\mathrm{(P1)}$ can be reformulated as
{\color{blue}
\begin{equation}
\label{deqn_ex43a}
\scalebox{0.85}{$
\begin{aligned}
\max_{\mathbf{p}_{tx}}\mathcal{F}(\mathbf{p}_{tx})=\max_{\mathbf{p}_{tx}}&\,\sum_{k=1}^{K}\rho\mathrm{R}_{k}+\left ( 1-\rho \right )\sum_{k=1}^{K}\sum_{i=1}^{3} \aleph_{i}\frac{1}{\bm{\mathrm{L}}_{k}^{ii}}\\
\mathrm{s.t.}\quad &{\rm(\ref{deqn_ex37A})},{\rm(\ref{deqn_ex37C})},
\end{aligned}$}
\end{equation}
where $\mathrm{R}_{k}=\sum_{n\in\mathcal{N}_{k}}\mathrm{log}\big(1+\alpha_{k}\Xi(\mathbf{f}_{k},\mathbf{s}_{k})p_{n}T_{e}/\eta_{0}\big)$, the corresponding terms of  $(\ref{deqn_ex61a})$ and $(\ref{deqn_ex62a})$ are replaced accordingly.}
To solve this non-convex problem, we find the optimal solution by improving the PGA algorithm in \cite{ref31}. 
Specifically, the $\mathbf{p}_{tx}^{(o)}$-based update rule is given by
{\color{blue}
\begin{equation}
\label{deqn_ex44a}
\scalebox{0.85}{$
\begin{aligned}
p_{tx,l}^{(o+1)}&=p_{tx,l}^{(o)}+\delta_{1}\left[\nabla_{\mathbf{p}_{tx}^{(o)}}\mathcal{F}\left(\mathbf{p}_{tx}^{(o)}\right)\right]_{l},\\
p_{tx,l}^{(o+1)}&=\mathcal{B}\left\{p_{tx,l}^{(o+1)}, D_{tx}, D_{min}^{tx}, D_{max}^{tx}\right\}.
\end{aligned}$}
\end{equation}
The first term represents the original update to $p_{tx,l}$ in the $(o+1)$-th iteration, and the projection function $\mathcal{B}\left \{ \cdot  \right \} $ ensures that the elements in each internal iteration do not exceed} the feasible region determined by constraints $\rm{(\ref{deqn_ex37A})}$ and $\rm{(\ref{deqn_ex37C})}$. 
In addition, $\nabla _{\mathbf{p}_{tx}^{(o)}}\mathcal{F}\left(\mathbf{p}_{tx}^{(o)}\right)$ represents the gradient of $\mathcal{F}\left(\mathbf{p}_{tx}^{(o)}\right)$ with respect to the variable $\mathbf{p}_{tx}^{(o)}$, and $\delta_{1}$ is the step size of the gradient ascent.

{\color{blue} Computing the gradient of $\mathcal{F}(\mathbf{p}_{tx}^{(o)})$ essentially involves taking the derivatives of the terms related to $\mathbf{p}_{tx}$, and then combining them through addition and multiplication.
By defining} $\mathbf{\Omega }_{1}^{k}=\mathrm{diag}(\big\{-\frac{2\pi cos\theta_{k}}{\lambda}sin(\frac{2\pi p_{tx,l}cos\theta_{k}}{\lambda})\big\}_{l=1}^{M_{tx}})$, $\mathbf{\Omega }_{2}^{k}=\mathrm{diag}(\big\{\frac{2\pi cos\theta_{k}}{\lambda}cos(\frac{2\pi p_{tx,l}cos\theta_{k}}{\lambda})\big\}_{l=1}^{M_{tx}})$, $\mathbf{\Omega }_{3}^{k}=\mathrm{diag}(\big\{\\cos(\frac{2\pi p_{tx,l}cos\theta_{k}}{\lambda})-\frac{2\pi cos\theta_{k}}{\lambda}p_{tx,l}sin(\frac{2\pi p_{tx,l}cos\theta_{k}}{\lambda})\big\}_{l=1}^{M_{tx}})$, $\mathbf{\Omega }_{4}^{k}=\mathrm{diag}(\big\{sin(\frac{2\pi p_{tx,l}cos\theta_{k}}{\lambda})+\frac{2\pi cos\theta_{k}}{\lambda}p_{tx,l}cos(\frac{2\pi p_{tx,l}cos\theta_{k}}{\lambda})\big\}_{l=1}^{M_{tx}})$, we can obtain
\begin{equation}
\label{deqn_ex45a}
\scalebox{0.82}{$
\begin{aligned}
\nabla _{{\color{blue}\mathbf{p}_{tx}}}\Xi\left(\mathbf{f}_{k},{\color{blue}\mathbf{s}}_{k}\right)&=2\mathbf{\Omega }_{1}^{k}({\color{blue}\mathbf{T}_{k}}\mathbf{f}_{k}+{\color{blue}\mathbf{D}}_{k}{\color{blue}\mathbf{s}}_{k})+2\mathbf{\Omega }_{2}^{k}({\color{blue}\mathbf{T}_{k}}{\color{blue}\mathbf{s}}_{k}-{\color{blue}\mathbf{D}}_{k}\mathbf{f}_{k}),\\
\nabla_{{\color{blue}\mathbf{p}_{tx}}}\Xi_{1}\left(\mathbf{f}_{k},{\color{blue}\mathbf{s}}_{k}, \mathbf{\Lambda}_{tx}\right)&=2\mathbf{\Omega}_{3}^{k}({\color{blue}\mathbf{T}_{k}}\mathbf{\Lambda}_{tx}\mathbf{f}_{k}+{\color{blue}\mathbf{D}}_{k}\mathbf{\Lambda}_{tx}{\color{blue}\mathbf{s}}_{k})\\&+2\mathbf{\Omega}_{4}^{k}({\color{blue}\mathbf{T}_{k}}\mathbf{\Lambda}_{tx}{\color{blue}\mathbf{s}}_{k}-{\color{blue}\mathbf{D}}_{k}\mathbf{\Lambda}_{tx}\mathbf{f}_{k}).
\end{aligned}$}
\end{equation}

The partial derivatives of $\Xi_{2}\left(\mathbf{f}_{k},{\color{blue}\mathbf{s}}_{k}, \mathbf{\Lambda}_{tx}\right)$ and $\Xi_{3}\left(\mathbf{f}_{k}\\,{\color{blue}\mathbf{s}}_{k}, \mathbf{\Lambda}_{tx}\right)$ can be calculated directly using the Jacobian.
\begingroup
\color{blue}
The projection function $\mathcal{B}(\cdot)$ \cite{ref32} of each MA $p_{tx,l}$ can be expressed as 
\begin{equation}
\label{deqn_ex46a}
max\big(p_{tx,l-1}+D_{sp}, min( p_{tx,l}^{(o+1)},D_{max}^{tx}-\left(M_{tx}-l \right)D_{sp})\big),
\end{equation}
where $p_{tx,0} =D_{max}^{tx}-D_{sp} $.
\endgroup
By iteratively updating {\color{blue}$p_{tx,l}^{(o+1)}$} using $(\ref{deqn_ex44a})$, 
the objective function in $(\ref{deqn_ex43a})$ converges to a constant value.

\subsection{Receive Antenna Design}
Given fixed vectors 
{\color{blue}$\mathbf{w}$, $\mathbf{p}$} and 
$\mathbf{p}_{tx}$, {\color{blue} only $\mathrm{LPCRLB}_{\hat{\theta}_{k}}$ related to the optimization variable $\mathbf{p}_{rx}$, $\mathrm{(P1)}$} can be simplified to the following maximized form, i.e.,
\begingroup
\color{blue}
\begin{equation}
\label{deqn_ex47a}
\scalebox{0.95}{$
\begin{aligned}
\max_{\mathbf{p}_{rx}}\quad& sin^{2}\theta_{k} \mathbf{p}^{T}\mathbf{g}_{kk}^{11} +\frac{4\nu_{k}sin^{2}\theta_{k}}{\lambda}(\mathbf{p}^{T}\mathbf{g}_{kk}^{31} +\frac{\nu_{k}}{\lambda}\mathbf{p}^{T}\mathbf{g}_{kk}^{33})+\bm{\mathcal{J}}_{kk}^{11},\\
\mathrm{s.t.}\quad& \quad\quad\quad\quad\quad\quad\quad\quad{\rm(\ref{deqn_ex37B})}, {\rm(\ref{deqn_ex37D})}.
\end{aligned}$}
\end{equation}
\endgroup
To effectively address this challenge, the PGA method is applied. 
Following (\ref{deqn_ex44a}), we first compute the derivative of the objective function with respect to $\mathbf{p}_{rx}$, then iteratively update {\color{blue}$p_{rx,ll}$} with a step size $\delta_{2}$, and finally enforce the constraints in ${\rm(\ref{deqn_ex37B})}$ and ${\rm(\ref{deqn_ex37D})}$.
Thus, the final objective function converges to a constant value. 
{\color{blue}
The gradient of the objective function depends on $\mathbf{g}_{kk}^{11}$ and $\mathbf{g}_{kk}^{31}$, where $\nabla_{\mathbf{p}_{rx}}\left[\mathbf{g}_{kk}^{11}\right]_{n} = (\frac{2\pi\gamma_{k}}{\lambda})^{2}\frac{Q}{\eta_{1}T_{e}}\big(2\mathbf{p}_{rx}\Xi(\mathbf{f}_{k},\mathbf{s}_{k})-\Xi_{2}(\mathbf{f}_{k},\mathbf{s}_{k},\mathbf{\Lambda}_{tx})-\Xi_{3}(\mathbf{f}_{k},\mathbf{s}_{k},\mathbf{\Lambda}_{tx})\big)$ and $\nabla_{\mathbf{p}_{rx}}\left[\mathbf{g}_{kk}^{31}\right]_{n} = \frac{Q(Q-1)(2\pi\gamma_{k})^{2}T_{s}}{2\lambda\eta_{1}T_{e}}\Xi(\mathbf{f}_{k},\mathbf{s}_{k})  $.}
Unlike the transmit antenna, the projection function $\mathcal{B} \left \{ {\color{blue}p_{rx,ll}^{(o+1)}}, D_{rx}, {D}_{min}^{rx},{D}_{max}^{rx}\right \} $ of the receive antenna obeys
\begingroup
\color{blue}
\begin{equation}
\label{deqn_ex48a}
max\big (D_{min}^{rx}+(M_{rx}-ll)D_{sp},min\big(p_{rx,ll}^{(o+1)}, p_{rx,ll+1}-D_{sp}\big)\big),
\end{equation}
where $p_{rx,M_{rx}+1}=D_{max}^{rx}+D_{sp}$.
\endgroup

{\color{blue}\subsection{Complexity Analysis and Scalability}}
The general PRE-SC-PGA algorithm for solving {\color{blue}$\mathrm{(P1)}$ } is presented in Algorithm 1. 
\begingroup
\color{blue}
The PRE-SC-PGA algorithm performs beamforming optimization, approximate rank-1 recovery, power vector optimization, and transceiver antenna position updates in each outer iteration.
In the beamforming stage, the semidefinite programmin (SDP) is solved using SCA and interior-point methods. Assuming $I_{bf}$ internal iterations, each SDP solution has a complexity of approximately 
$\mathcal{O}\big(K^{3}M_{tx}^{6})$, resulting in a total beamforming complexity of 
$\mathcal{O}\big(I_{bf}K^{3}M_{tx}^{6})$.
The power vector optimization is a fractional quadratic problem with complexity 
$\mathcal{O}\big(K)$, which is negligible. 
The antenna position update employs PGA, whose complexity grows quadratically with 
$M_{tx}$ or $M_{tx}^{2}$, contributing little to the overall cost. 
Therefore, the total complexity over $I_{out}$ outer iterations is $\mathcal{O}\big(I_{out}I_{bf}K^{3}M_{tx}^{6})$.
Hence, the complexity of the proposed algorithm grows with the sixth power of the number of the transmit antennas and with the third power of the number of vehicles.
Under the parameter settings in this paper, the algorithm takes three hundred seconds to run.
As the number of transmit antennas or vehicles increases, the computational cost increases rapidly, low-rank approximations can be employed to enhance the scalability of large-scale MIMO systems.
\endgroup

\begin{algorithm}[!t]
\captionsetup{font=footnotesize} 
\footnotesize
{\color{blue}
\caption{PRE-SC-PGA for Solving $\mathrm{(P1)}$.}\label{alg:alg1}
\begin{algorithmic}
\STATE 
\STATE {$\textbf{Initialize}$ the transmit power vector as $\mathbf{p}^{0}$, MA positions as $\mathbf{p}_{tx}^{0}$ and 
$\mathbf{p}_{rx}^{0}$};\\
\STATE Introduce quantization factors  $\left \{\aleph_{i}\right \}_{i=1}^{3} $ to reconstruct $\mathrm{(P1)}$;
\STATE $\textbf{Repeat}$
\STATE \hspace{0.2cm} Update $\hat{\mathbf{w}}^{*}$ by solving $\mathrm{(P1.2)}$ using CVX solver \cite{ref30} and Gaussian \\\hspace{0.3cm}randomization;
\STATE \hspace{0.2cm} Update $\mathbf{p}$ by solving (P1);
\STATE \hspace{0.3cm}$\textbf{Repeat}$
\STATE \hspace{0.6cm}Alternately update $p_{tx,l}^{(o+1)}$ using the rule of (\ref{deqn_ex42a}) with (\ref{deqn_ex44a});
\STATE \hspace{0.3cm}$\textbf{Until}$ Converges
\STATE \hspace{0.3cm}$\textbf{Repeat}$
\STATE \hspace{0.6cm}Alternately update $p_{rx,ll}^{(o+1)}$ using the rule of (\ref{deqn_ex42a}) with (\ref{deqn_ex46a});
\STATE \hspace{0.3cm}$\textbf{Until}$ Converges
\STATE $\textbf{Until}$ The objective of $\mathrm{(P1)}$ converges to a prescribed accuracy.
\end{algorithmic}
\label{alg1}}
\end{algorithm}

\begingroup
\color{blue}
\section{RPDPSO Algorithm Design For Sum-Rate Maximization Problem}
Because the optimization problem focuses solely on determining the transmit-antenna positions, the resulting search space is relatively limited for PSO-type methods. 
While PSO generally incurs higher computational cost than AO, the required runtime remains acceptable and PSO provides superior rate performance, primarily due to its lower susceptibility to local optima. 
In light of the need for both high communication performance and accurate sensing in dynamic vehicular scenarios, we propose a heuristic algorithm, referred to as RPDPSO, which maximizes the communication sum-rate subject to the sensing LPCRLB constraints.

\subsection{RPDPSO Algorithm Framework Design}
This section focuses on research in practical dynamic scenarios.
On the one hand, it is necessary to predict channel conditions more favorable for vehicular communications in the $m$-th time slot based on the estimated parameters from the $(m-1)$-th time slot and pre-deploying the transmit antenna positions.
On the other hand, real-time beam calibration needs to be performed during the $m$-th time slot.
As shown in the Fig.\ref{fig4}, we designed a two-stage algorithm framework.
In the $(m-1)$-th time slot, the proposed heuristic algorithm is utilized to obtain the optimal transmit antenna positions, while in the $m$-th time slot, the transmit power and beamforming vectors are re-optimized based on the currently estimated channel to enhance the communication sum-rate.
Note that the calculated PCRLB at this time is equivalent to the updated MSE matrix in EKF prediction.
Furthermore, using the currently received sensing echo signal and the predicted useful sensing echo signal, state tracking can be performed to obtain $\hat{\bm{\zeta}}_{m}$. 
Combining state tracking and the updated MSE matrix, the motion parameters in the $(m+1)$-th time slot can be predicted.
The same optimization and real-time beam alignment are then performed sequentially to achieve a cross-time slot communication-sensing joint adaptive optimization closed loop.

\begin{figure}
\hspace{-0.4cm}
\includegraphics[width=3.7in]{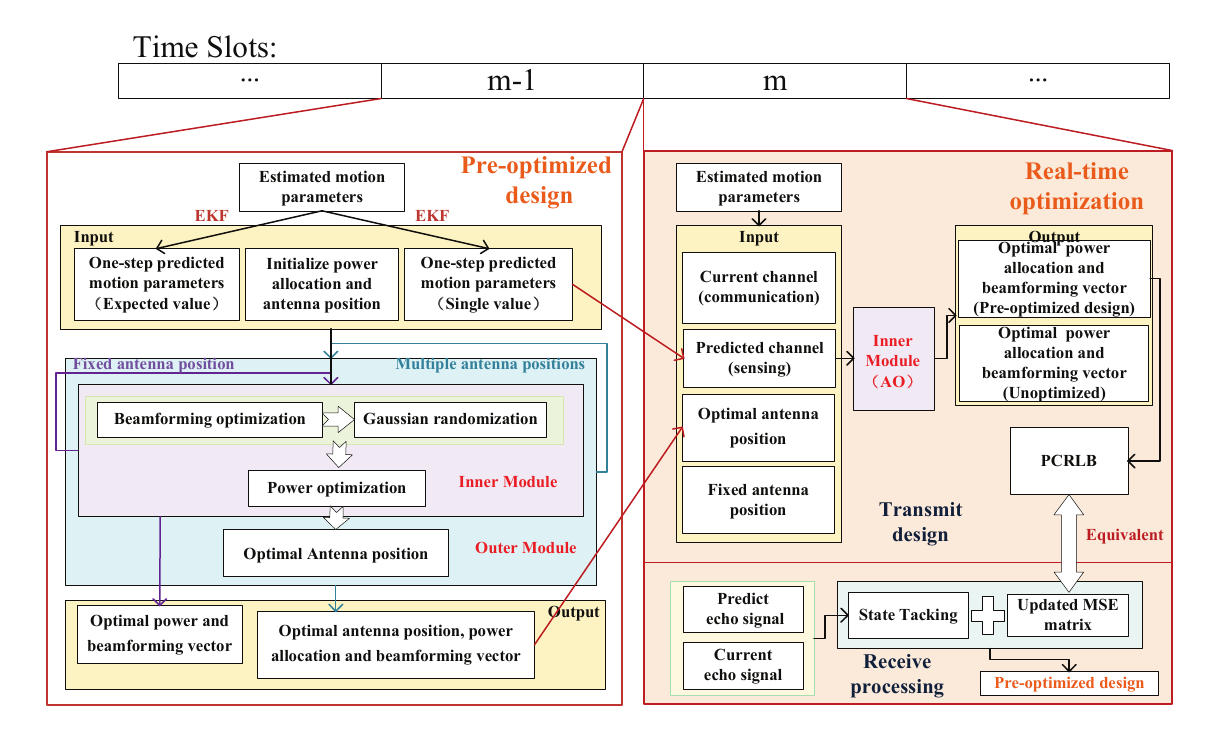}
    \caption{\underline{\textbf{This figure is newly added}} The RPDPSO algprithm framework for sum-rate maximization.}
\label{fig4}
\end{figure}

\subsection{Beamforming Design}
The beamforming design in this section is similar to the algorithm presented in Section IV-C, except that the objective function focuses solely on communication sum-rate, while the LPCRLBs of the DoD, distance, and velocity are assumed to be known. 
Accordingly, the problem can be formulated as
\begin{subequations}
\begin{align}
&\mScale{\mathrm{(P2.1)}\quad \max_{\mathbf{W}, \boldsymbol{\kappa}, \boldsymbol{\epsilon},\boldsymbol{\varepsilon}}\,\,\, \mathcal{F}_{1}(\mathbf{W})}\nonumber\\
\mathrm{s.t.}\,\,
&\mScale{(\rm{\ref{deqn_ex41A}})-(\rm{\ref{deqn_ex41C}}) ,}\nonumber\\
&\mScale{{\color{blue}\bm{\mathcal{Q}}_{kk}^{11}+\bm{\mathcal{J}}_{kk}^{11}>=\frac{1}{\varsigma_{k}^{\theta}},}}\label{deqn_ex49A}\\
&\mScale{{\color{blue}\begin{bmatrix}\frac{4}{c^{2}}\mathbf{p}^{T}\mathbf{g}_{kk}^{22}+\bm{\mathcal{J}}_{kk}^{22}-\frac{1}{\varsigma_{k}^{d}} & \frac{4cos\theta_{k}}{c\lambda}\mathbf{p}^{T}\mathbf{g}_{kk}^{23} +\bm{\mathcal{J}}_{kk}^{23}\\\frac{4cos\theta_{k}}{c\lambda}\mathbf{p}^{T}\mathbf{g}_{kk}^{32} +\bm{\mathcal{J}}_{kk}^{32}  &\frac{4cos^{2}\theta_{k}}{\lambda^{2}}\mathbf{p}^{T}\mathbf{g}_{kk}^{33}+\bm{\mathcal{J}}_{kk}^{33}
    \end{bmatrix} \succeq 0,}}\label{deqn_ex49B}\\
&\mScale{{\color{blue}\begin{bmatrix}\frac{4cos^{2}\theta_{k}}{\lambda^{2}}\mathbf{p}^{T}\mathbf{g}_{kk}^{33}+\bm{\mathcal{J}}_{kk}^{33}-\frac{1}{\varsigma_{k}^{\nu}} & \frac{4cos\theta_{k}}{c\lambda}\mathbf{p}^{T}\mathbf{g}_{kk}^{23} +\bm{\mathcal{J}}_{kk}^{23}\\\frac{4cos\theta_{k}}{c\lambda}\mathbf{p}^{T}\mathbf{g}_{kk}^{32} +\bm{\mathcal{J}}_{kk}^{32}  & \frac{4}{c^{2}}\mathbf{p}^{T}\mathbf{g}_{kk}^{22}+\bm{\mathcal{J}}_{kk}^{22}
    \end{bmatrix} \succeq 0,}\label{deqn_ex49C}}
\end{align}
\end{subequations}
After solving for $\mathbf{W}_{n}$ by ignoring the rank-one constraint, the optimal solution $\hat{\mathbf{w}}_{n}^{\ast}$ is obtained by Gaussian randomization.

\subsection{Transmit Power Design}
For $\mathrm{(P2)}$, constrains $\rm(\ref{deqn_ex38D})$ and $\rm(\ref{deqn_ex38E})$ are not convex functions relative to the variable $\mathbf{p}$.
Fortunately, we can transform these constraints into convex constraints and then construct a SDP problem for optimization. 
Concretely, looking back at (\ref{deqn_ex31a}), it can be seen that $\mathrm{LPCRLB}_{\hat{d}_{k}}$ is the $(1,1)$-th element of $\mathrm{LPCRLB}_{k}^{d\nu}$, which can be represented as
\begin{equation}
\label{deqn_ex50a}
\mathrm{LPCRLB}_{\hat{d}_{k}}=\frac{\bm{\mathcal{Q}}_{kk}^{33}+\bm{\mathcal{J}}_{kk}^{33}}{(\bm{\mathcal{Q}}_{kk}^{22}+\bm{\mathcal{J}}_{kk}^{22})(\bm{\mathcal{Q}}_{kk}^{33}+\bm{\mathcal{J}}_{kk}^{33})-(\bm{\mathcal{Q}}_{kk}^{23}+\bm{\mathcal{J}}_{kk}^{23})^{2}}.
\end{equation}
Considering that the value is less than or equal to $\varsigma_{k}^{d}$, the constraint $\rm(\ref{deqn_ex50a})$ can be equivalent to $\big(\varsigma_{k}^{d}(\bm{\mathcal{Q}}_{kk}^{22}+\bm{\mathcal{J}}_{kk}^{22})-1\big)(\bm{\mathcal{Q}}_{kk}^{33}+\bm{\mathcal{J}}_{kk}^{33})-\varsigma_{k}^{d}(\bm{\mathcal{Q}}_{kk}^{23}+\bm{\mathcal{J}}_{kk}^{23})^{2}\ge0$. 
By constructing the inequality into a matrix and following 
\begin{equation}
\label{deqn_ex51a}
   \mathbf{\Psi }_{k}^{d}=\left [ \begin{matrix}\bm{\mathcal{Q}}_{kk}^{33}+\bm{\mathcal{J}}_{kk}^{33}& \sqrt{\varsigma_{k}^{d}}(\bm{\mathcal{Q}}_{kk}^{23}+\bm{\mathcal{J}}_{kk}^{23})\\
  \sqrt{\varsigma_{k}^{d}}(\bm{\mathcal{Q}}_{kk}^{32}+\bm{\mathcal{J}}_{kk}^{32} ) &\varsigma_{k}^{d}(\bm{\mathcal{Q}}_{kk}^{22}+\bm{\mathcal{J}}_{kk,}^{22})-1
\end{matrix} \right ], 
\end{equation}
we can get $\mathbf{\Psi }_{k}^{d}\succeq0 $.
Similarly, the velocity constraint can be rewritten as $\mathbf{\Psi }_{k}^{\nu}\succeq0 $.
Based on the transformation of the above constraints, we can reformulate $\mathrm{(P2)}$ as
\begin{equation}
\begin{aligned}
\label{deqn_ex52a}
\mathrm{(P2.2)}\quad &\min_{\mathbf{p}} \quad -\sum_{k=1}^{K}\mathrm{R}_{k}  
\\& \mathrm{s.t.}\quad {\rm(\ref{deqn_ex38C})},\mathbf{\Psi }_{k}^{d}\succeq 0, \mathbf{\Psi }_{k}^{\nu}\succeq 0, {\rm(\ref{deqn_ex38F})}.  
\end{aligned}
\end{equation}
Obviously, $\mathrm{(P2.2)}$ is a convex optimization problem and can be solved by CVX toolbox \cite{ref30}.
In addition, for the optimization problem $\mathrm{(P2)}$ with nonlinear constraints, the Lagrange multiplier method can be used to equivalently transform it into an unconstrained problem for solution. 
The closed-form solution for the power allocated to each subcarrier is $p_{n}= max(0,\frac{1}{(\lambda_{1}-\jmath )ln2}-\frac{\eta_{0}}{\left|\mathbf{a}^{H}(\mathbf{p}_{tx},\theta_{k})\mathbf{w}_{n}\right|^2T_{e}})$, where $\lambda_{1}$ and $\jmath$ are Lagrange multipliers related to the transmit power and the sensing LPCRLBs, respectively.

\subsection{Design of RPDPSO Algorithm}
To address the strong coupling between the power allocation vector and the beamforming vector, while simultaneously satisfying the antenna position constraint, we propose a RPDPSO algorithm. 
The details are as follows.

Given the dimension of the number of particles as $M_{tx}$, define the number of particles as $N_{\bar{p}}$, the velocity and position of each particle can be initialized as $\mathbf{v}_{\bar{p}}^{(0)}\in \mathbb{R}^{M_{tx}\times1}$ and $\bar{\mathbf{u}}_{\bar{p}}^{(0)}\in \mathbb{R}^{M_{tx}\times1}$, respectively.
For the proposed algorithm, the position of the transmit antenna is treated as a known value as input, which satisfies the antenna constraint {\rm(\ref{deqn_ex38B})}. 
The only remaining optimization variables are the power allocation vector and the beamforming vector.

\textit{1) Define Fitness Function:}
Taking advantage of the concept of the Lagrange penalty term, a penalty function $\mathcal{P}(\cdot) = \imath\delta(\cdot)$ with the penalty factor $\imath$ is introduced to calculate the number of pairs of antenna positions in each particle that violate the minimum distance constraint {\rm(\ref{deqn_ex38A})}.
Define $\mathcal{S}(\cdot)$ as the optimal communication sum-rate obtained after optimizing the beamforming vector and the power allocation vector for each particle, the fitness function can be expressed as
\begin{equation}
\label{deqn_ex53a}
\mathcal{F}(\bar{\mathbf{u}}_{\bar{p}}^{(ir)}) = \mathcal{S}(\bar{\mathbf{u}}_{\bar{p}}^{(ir)}) + \mathcal{P}(\bar{\mathbf{u}}_{\bar{p}}^{(ir)}),
\end{equation}
where $ir\in\left\{1,2,\cdots,\mathrm{Iter}\right\}$ represents the $ir$-th iteration of the particles and the value of $\mathcal{S}(\bar{\mathbf{u}}_{\bar{p}}^{(ir)})$ is negative.

\textit{2) Reflection-Projection Boundary Handling:}
The fitness value of each particle is computed according to {\rm(\ref{deqn_ex53a})}, and each particle's current position and corresponding fitness value are initialized as its personal best position and personal best fitness.
Among all particles, the particle with the lowest fitness and its corresponding position is selected as the initial global best, serving to guide the subsequent iterative search.
For the $ir$-th iteration, the velocity and position update criteria for each particle are as follows
\begin{equation}
\label{deqn_ex54a}
\scalebox{0.75}{$
\begin{aligned}
\mathbf{v}_{\bar{p}}^{(ir)} &=w^{(ir)}\mathbf{v}_{\bar{p}}^{(ir-1)} + \mathit{c}_{1}\mathbf{e}_{1} \odot(\bar{\mathbf{u}}_{pbest,\bar{p}}-\bar{\mathbf{u}}_{\bar{p}}^{(ir-1)}) +\mathit{c}_{2}\mathbf{e}_{2} \odot(\bar{\mathbf{u}}_{gbest,\bar{p}}-\bar{\mathbf{u}}_{\bar{p}}^{(ir-1)}),\\
\mathbf{v}_{\bar{p}}^{(ir)} &= \mathcal{B}\left\{ \mathbf{v}_{\bar{p}}^{(ir)}\right\}, \quad \bar{\mathbf{u}}_{\bar{p}}^{(ir)}  = \mathcal{B}\left\{ \bar{\mathbf{u}}_{\bar{p}}^{(ir-1)}+\mathbf{v}_{\bar{p}}^{(ir)}\right\},\\
\mathbf{v}_{\bar{p}}^{(ir)} &=\mathbf{v}_{\bar{p}}^{(ir)} -(1+\mathit{s}_{f}^{2})\big([\mathbf{v}_{\bar{p}}^{(ir)}\cdot \mathbf{1}_{(\bar{\mathbf{u}}_{\bar{p}}^{(ir)}>D_{max}^{tx})}]-[\mathbf{v}_{\bar{p}}^{(ir)} \cdot\mathbf{1}_{(\bar{\mathbf{u}}_{\bar{p}}^{(ir)}<D_{min}^{tx})}]\big),
\end{aligned}$}
\end{equation}
where $w^{(ir)}$ is a linear inertia weight function with $w = w_{max}-\frac{ir}{Iter}(w_{max}-w_{min})$. 
$\mathit{c}_{1}$ and $\mathit{c}_{2}$ are the personal and global learning factors, respectively. 
$\mathbf{e}_{1}$ and $\mathbf{e}_{2}$ are random vectors, whose terms follow a uniform distribution within $[0,1]$. 
$\mathcal{B}\left\{ \mathbf{v}_{\bar{p}}\right\} = \mathrm{max}(-\mathbf{v}_{max},\mathrm{min}(\mathbf{v}_{\bar{p}}, \mathbf{v}_{max})$ can suppress excessively large step sizes, preventing particles from jumping out of the feasible region in a single update and reducing oscillations.
$\mathbf{v}_{max}=\mathit{s}_{f}^{1}(D_{max}^{tx}-D_{min}^{tx})$ with the scaling factor $\mathit{s}_{f}^{1}\in [0,1]$.
$\mathit{s}_{f}^{2}$ is also a scaling factor, which characterizes the reflection attenuation coefficient.
$\mathbf{1}_{(\cdot)}$ is an indicator function that takes the value 1 when the condition is true and 0 otherwise.

\textit{3) Spatial-Aware Dynamic Pruning:}
After each iteration of the particle position updates, the fitness value of each particle is recalculated.
If the fitness value of the particle exceeds its current personal best, the particle's personal best position and fitness are updated accordingly.
Moreover, the global optimal position and fitness value are only updated when the fitness value of the particle exceeds the current global optimum, and the corresponding metrics are recorded.

To avoid wasting computational resources on particle in low-yield regions, we first define the spatial distance between a particle and the global optimum, which can be expressed as
\begin{equation}
\label{deqn_ex55a}
\begin{aligned}
D_{\bar{p}} = \left \|\bar{\mathbf{u}}_{\bar{p}}^{(ir)} -\bar{\mathbf{u}}_{gbest,\bar{p}} \right \| _{2}.
\end{aligned}
\end{equation}
Pruning is performed by introducing a threshold $T_{\bar{p}}=\mathit{s}_{f}^{3}\left \|D_{max}^{tx}-D_{min}^{tx}\right \|$, which yeilds
\begin{equation}
\label{deqn_ex56a}
\begin{aligned}
\mathcal{A}^{(ir)}=\left\{\bar{p}:D_{\bar{p}}^{(ir)}>T_{\bar{p}}\right\}.
\end{aligned}
\end{equation}
Excessive pruning may lead to population collapse.
To address this, we design an adaptive retention mechanism.
When the number of particles exceeds a predefined threshold, a pruning operation is performed to remove redundant particles. 
Conversely, when the particle count falls below the threshold, new particles are generated through random sampling based on the retained optimal particles to maintain population diversity.

\subsection{Convergence and Complexity Analysis}
As shown in Algorithm 2, its convergence performance of the overall algorithm depends on the internal beamforming and power allocation algorithm, as well as the external RPDPSO algorithm.
The convergence of the beamforming algorithm can be derived by referring to the proof in \textbf{Appendix B}.
The power allocation problem is a convex optimization problem and satisfies the water-filling theorem under total power and nonnegative power constraints, possessing a unique closed-form optimal solution. 
Therefore, convergence is guaranteed.
For each iteration, only the position with smaller fitness value is selected as the new global best position.
Therefore, the global best fitness value is monotonically non-increasing during the iterative process. 
Moreover, the negative value of the communication sum-rate is lower-bounded. 
Consequently, the convergence of the RPDPSO  algorithm is guaranteed.
The complexities of the optimization algorithm and the RPDPSO algorithm are $\mathcal{O}(I_{bf}K^{3}M_{tx}^{6})$ and $\mathcal{O}(\sum_{k=1}^{K}\sum_{ir=1}^{\mathrm{Iter}}P_{act}^{ir})$, respectively. 
Here, $P_{act}^{ir}$ is the number of active particles in the $ir$-th iteration.
Therefore, the overall complexity is $\mathcal{O}(I_{bf}K^{3}M_{tx}^{6}\sum_{k=1}^{K}\sum_{ir=1}^{\mathrm{Iter}}P_{
act}^{ir})$.
\endgroup

\begin{algorithm}[!t]
\captionsetup{font=footnotesize} 
\footnotesize
\caption{RPDPSO for Solving $\mathrm{(P2)}$}\label{alg:alg2}
\begin{algorithmic}[1]
{\color{blue}
\STATE \textbf{Input:} $\bm{\zeta}, M_{tx}, N_{\bar{p}}, c_{1}, c_{2}, \mathbf{e}_{1}, \mathbf{e}_{2}, w_{\min}, w_{\max}, \mathit{s}_{f}^{1},\mathit{s}_{f}^{2}, \mathit{s}_{f}^{3}, T_{\bar{p}}, \mathbf{p}, \mathrm{Iter},$ \\\quad\quad\quad$\imath, D_{\min}^{tx}, D_{\max}^{tx}, N_{th}$.
\STATE \textbf{Output:} $\bar{\mathbf{u}}_{gbest}$, $\mathbf{p}_{opt}$, $\mathbf{w}_{opt}$.
\STATE Initialize the velocity and position of each particle as $\mathbf{v}_{\bar{p}}^{(0)}$ and $\bar{\mathbf{u}}_{\bar{p}}^{(0)}$, respectively.
\STATE Evaluate the fitness value for each particle using (\ref{deqn_ex51a}).
\STATE Obtain the personal best position $\bar{\mathbf{u}}_{pbest,\bar{p}} = \bar{\mathbf{u}}_{\bar{p}}^{(0)}$ and the global best position 
$\bar{\mathbf{u}}_{gbest,\bar{p}} = \arg\min_{\bar{\mathbf{u}}_{\bar{p}}^{(0)}} \left\{\mathcal{F}(\bar{\mathbf{u}}_{1}^{(0)}), \cdots, \mathcal{F}(\bar{\mathbf{u}}_{N_{\bar{p}}}^{(0)})\right\}.$
\FOR{$ir = 1: \mathrm{Iter}$}
  \STATE Update the inertia weight $w^{(ir)}$ and the number of active particles $P_{act}^{(ir)}$.
  \FOR{$\bar{p} = 1: N_{\bar{p}}$}
    \STATE Update the velocity and position of the $\bar{p}$-th particle according to (\ref{deqn_ex54a}).
    \STATE Calculate the fitness value $\mathcal{F}(\bar{\mathbf{u}}_{\bar{p}}^{(ir)})$ by solving the beamforming and power allocation subproblems.
    \IF{$\mathcal{F}(\bar{\mathbf{u}}_{\bar{p}}^{(ir)}) < \mathcal{F}(\bar{\mathbf{u}}_{pbest,\bar{p}}^{(ir)})$}
      \STATE Update $\bar{\mathbf{u}}_{pbest,\bar{p}}^{(ir)} = \bar{\mathbf{u}}_{\bar{p}}^{(ir)}$.
    \ENDIF
    \IF{$\mathcal{F}(\bar{\mathbf{u}}_{\bar{p}}^{(ir)}) < \mathcal{F}(\bar{\mathbf{u}}_{gbest,\bar{p}}^{(ir)})$}
      \STATE Update $\bar{\mathbf{u}}_{gbest,\bar{p}}^{(ir)} = \bar{\mathbf{u}}_{\bar{p}}^{(ir)}$.
    \ENDIF
  \ENDFOR
  \IF{$P_{act}^{(ir)} > N_{th}$}
    \STATE Prune particles according to (\ref{deqn_ex56a}), retaining high-yield particles as active ones.
    \ENDIF
  \IF{$P_{act}^{(ir)} < N_{th}$}
    \STATE Preserve the best particles and generate new ones through random sampling.
  \ENDIF
\ENDFOR}
\end{algorithmic}
\end{algorithm}

\section{Simulation Results}
This section evaluates the performance of the MA-assisted V2I system through extensive simulations. We first present the simulation parameters and baseline configurations. For the case without sensing constraints, we analyze the convergence behavior of the PRE-SC-PGA algorithm and assess its performance with respect to the number of antennas, transmit power, and feasible region size. We then examine its performance trade-offs compared with baseline methods. For practical scenarios with sensing constraints, we further evaluate the performance gains brought by the pre-optimized design and analyze the impact of the optimal antenna positions on the communication and sensing performance in the current slot.

\subsection{Simulation Settings}

We consider a mmWave MIMO-OFDM system operating at $28$GHz carrier frequency and $100$MHz bandwidth. 
Active subcarriers and OFDM symbols in the system are set to {\color{blue}$N=32$, $Q=7$}. 
According to the fifth generation New Radio standard \cite{ref33}, the subcarrier spacing is set to $\triangle f=120$kHz and the symbol duration is set to $T_{s}=8.92\mu s$. 
Assume that the transmit antennas $M_{tx}=8$ and the receive antennas $M_{rx}=8$ are set in the system to serve the {\color{blue}$K=2$} single-antenna vehicles and consider a one-dimensional line segment with a variable length of {\color{blue}$D_{max}=D_{max}^{rx}-D_{min}^{rx}=D_{max}^{tx}-D_{min}^{tx}$} for performance analysis. 
The minimum distance between two adjacent MA antennas is set as {\color{blue}$D_{sp}=\lambda/2$}. 
The antenna spacing between the transmit and receive antennas is not less than {\color{blue}$D_{tr}=\lambda/2$}.
The initial position parameters of two vehicles are set as {\color{blue}$\left \{ \theta_{k} \right \}_{k=1}^{2}=\left \{ 9.2^{\circ},12^{\circ} \right \}$, $\left \{ d_{k} \right \}_{k=1}^{2}=\left \{400,410\right \} m$ and $\left \{ \nu_{k} \right \}_{k=1}^{2}=\left \{20,18\right \} m/s$.} 
The total transmission power is set to 1$W$. 
The power spectral densities of the communication receiver and the radar receiver are $\eta_{0}=1e-23$, $\eta_{1}=1.1e-25$. 
The attenuation coefficient is $\beta_{k}=\sqrt{\frac{\lambda ^{2}RCS }{\left ( 4\pi \right )^{3}\left (d_{k}/2 \right ) ^{4}} } $, where $RCS = 0.1$. The path loss at the reference distance $d_{0}=1m$ is $\alpha_{0}=-70 dB$, and the path loss exponent is $\zeta = 2.55$. 
{\color{blue}
To evaluate the overall performance of the proposed algorithm, we adopt the following three schemes as benchmarks for comparison.

\textbf{Uniform linear array (ULA) with half-wavelength antenna
spacing (ULAH) \cite{ref14}:} $\mathbf{p}_{tx}$ and $\mathbf{p}_{rx}$ are set according to the ULA, with antenna spacing of half-wavelength.

\textbf{Search-based projected gradient ascent (SPGA) \cite{ref18}:}  It includes three stages: i) initial point search, ii) gradient ascent updating, and iii) feasibility region projection. 

\textbf{Dynamic neighborhood
 pruning particle swarm optimization (DNPPSO) \cite{ref34}:} A two-loop dynamic neighborhood pruning PSO algorithm.
 In the outer loop, the antenna positions are updated by evaluating the fitness function, while in the inner loop, beamforming are optimized for each candidate position to compute its corresponding fitness value.}

\subsection{Convergence performance of Proposed PRE-SC-PGA Algorithm}
\begin{figure}
\centering
\includegraphics[width=3.5in]{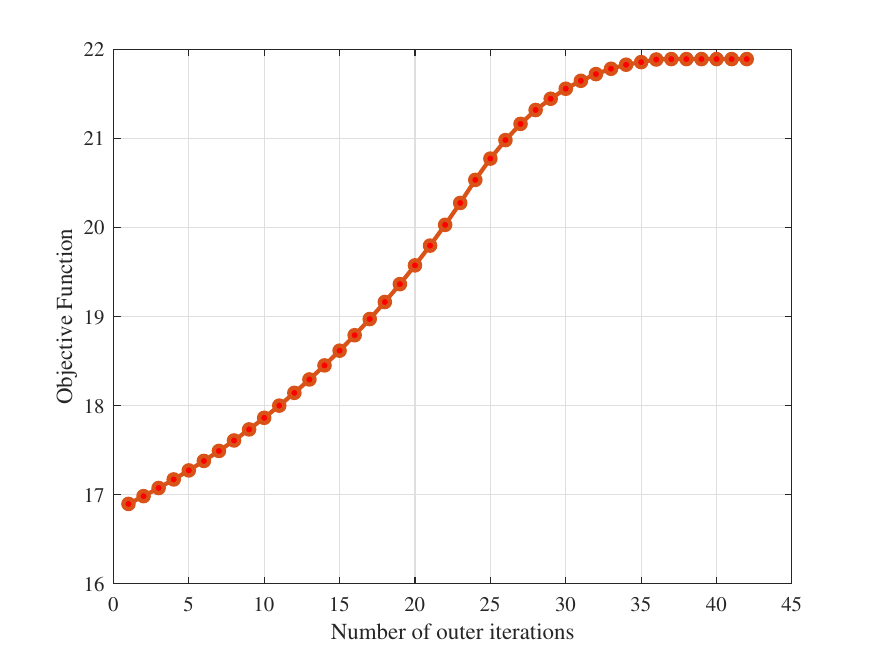}
\caption{\underline{\textbf{This figure is updated}}Convergence behavior of the PRE-SC-PGA algorithm.}
\label{fig5}
\end{figure}

Fig. \ref{fig5} illustrates the general convergence behavior of the {\color{blue} proposed PRE-SC-PGA algorithm for solving the optimization problems (P1)} when the feasible region is {\color{blue}$9\lambda$}.
As can be observed, the parameter variables $\mathbf{w}$, $\mathbf{p}_{tx}$, {\color{blue}$\mathbf{p}$} and $\mathbf{p}_{rx}$ fully interacted with each other until the objective function converged to a prescribed accuracy within around {\color{blue}50} outer iterations, thus validating the effectiveness of the proposed algorithm in achieving variable
coordination and optimization effects.

\subsection{Performance of PRE-SC-PGA Algorithm}

\begin{figure}[!t]
\centering
\includegraphics[width=3.5in]{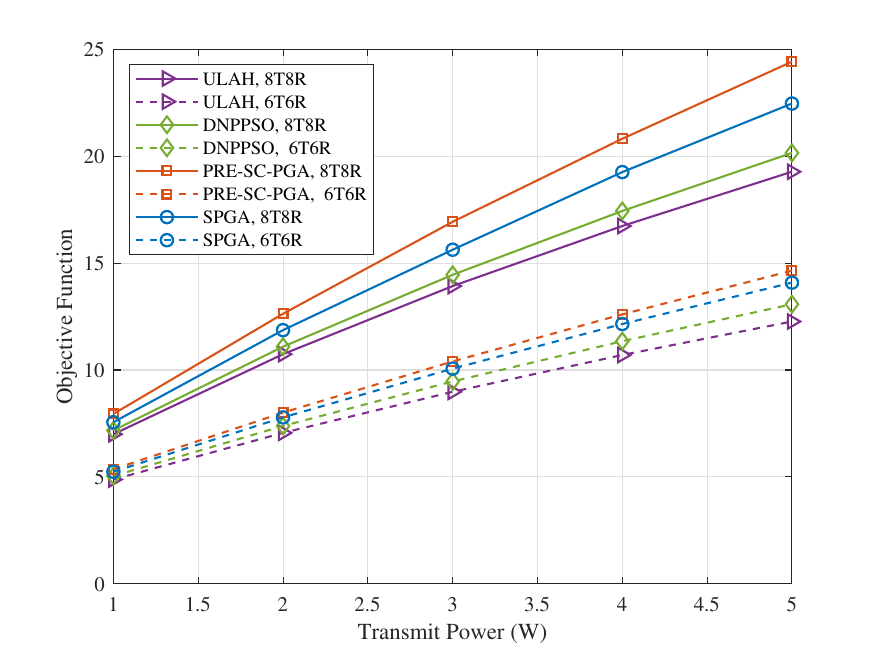}
\caption{\underline{\textbf{This figure is updated}}The weighted-sum of communication and sensing performance versus the total transmit power $P_{T}$ with $\rho=0.5$, $D_{max}={\color{blue}7}\lambda$.}
\label{fig6}
\end{figure}

Fig. \ref{fig6} shows that the weighted-sum of sum-rate and sensing performance of {\color{blue}three} types of motion parameters vary with the total transmit power when the feasible region size is {\color{blue}7}$\lambda$ and the weighting factor is 0.5 under 8-transmit 8-receive antennas. 
In fact, the increase in transmit power is equivalent to an improvement in the average SNR, which further reflects an increase in the energy of the received communication signal and the echo signal. 
Therefore, the performance of the weighted-sum achieved by the beamforming {\color{blue}design and power allocation for both the benchmark schemes and the proposed scheme} is improved with increasing transmit power.
For 8-transmit 8-receive antennas, the weighted-sum performance achieved by the proposed {\color{blue}PRE-SC-PGA  algorithm is significantly better than the other three beamforming algorithms.} 

\begin{figure}[!t]
\centering
\includegraphics[width=3.5in]{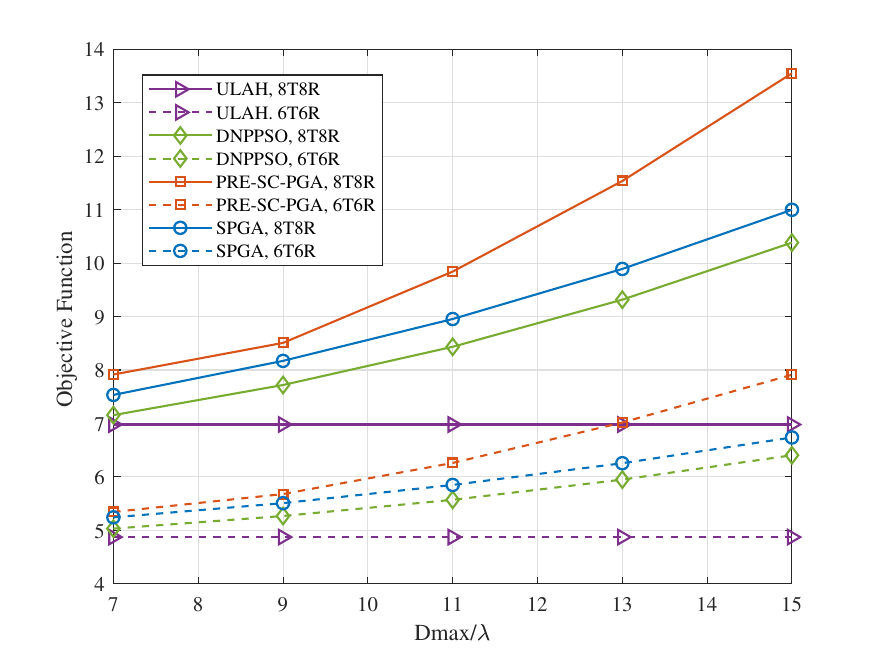}
\caption{\underline{\textbf{This figure is updated}}The weighted-sum of communication and sensing performance versus the size of feasible region with $\rho=0.5$, $P=1W$.}
\label{fig7}
\end{figure}

Fig. \ref{fig7} illustrates the performance of the weighted-sum of communication and sensing for {\color{blue}all schemes} under 8-transmit 8-receive antennas and {\color{blue}6-transmit 6-receive} antennas, respectively, when the feasible region size changes from {\color{blue}7$\lambda$ to 15$\lambda$}. 
As the feasible region increases, {\color{blue}all the MA-assisted schemes} can obtain more DoF to improve channel conditions, resulting in an improvement in their objective function values.
{\color{blue}In this context, the proposed PRE-SC-PGA algorithm attains optimal performance among all considered algorithms.}
Actually, {\color{blue}ULAH is the} special case of the proposed MA-assisted scheme and its antenna position does not change with the size of the feasible region, thus the objective function value always remains unchanged.
It is shown that when the feasible region size is {\color{blue}13}$\lambda$, the objective function values obtained using {\color{blue}the proposed PRE-SC-PGA algorithm under 6-transmit 6-receive} antennas have exceeded those obtained using {\color{blue}ULAH} with 8-transmit 8-receive antenna.
The result suggests that with fewer antennas and proper deployment within a feasible region, it is possible to achieve a weighted-sum performance comparable to {\color{blue}ULAH} with more antennas. 
This highlights the potential of {\color{blue}the proposed algorithm} to maintain excellent performance while substantially reducing hardware costs.

\subsection{Trade-off Between Communication and Sensing Performance}

\begin{figure}[!t]
\centering
\includegraphics[width=3.5in]{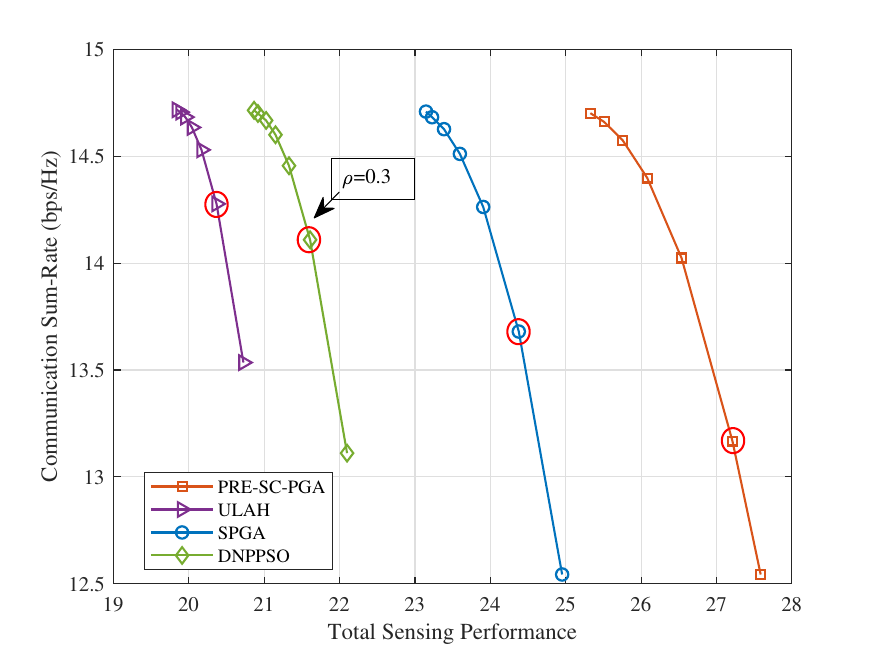}
\caption{\underline{\textbf{This figure is updated}} The trade-off between communication and total sensing performance {\color{blue}with $P=1W$, $D_{max}=7\lambda$.}}
\label{fig8}
\end{figure}

As illustrated in {\color{blue}Fig. \ref{fig8}}, we plot the relationship between the communication performance and the overall sensing performance {\color{blue}under four different beamforming schemes to examine the benefits brought by the MA assistance.}
The overall sensing performance {\color{blue}is defined as} the weighted-sum of the inverse of the LPCRLBs corresponding to the {\color{blue}three motion parameters}, where a larger value indicates better {\color{blue}sensing accuracy}. 
Obviously, {\color{blue}all schemes exhibit a trade-off between communication performance and sensing performance, where improving the communication performance inevitably degrades the sensing performance. }
It is worth highlighting that the performance change of the trade-off curve from left to right symbolizes the performance change of the trade-off factor from 1 to 0. 
{\color{blue}
The red circles in the figure highlight the communication and sensing performance achieved by each scheme when the weighting factor is set to 0.3. 
The marked results demonstrate that MA assistance can substantially enhance the sensing performance, albeit at the cost of some communication performance. 
Furthermore, for the same communication sum-rate, the proposed PRE-SC-SCA algorithm achieves the best sensing performance among all evaluated algorithms.}

\begin{figure}[!t]
\centering
\includegraphics[width=3.5in]{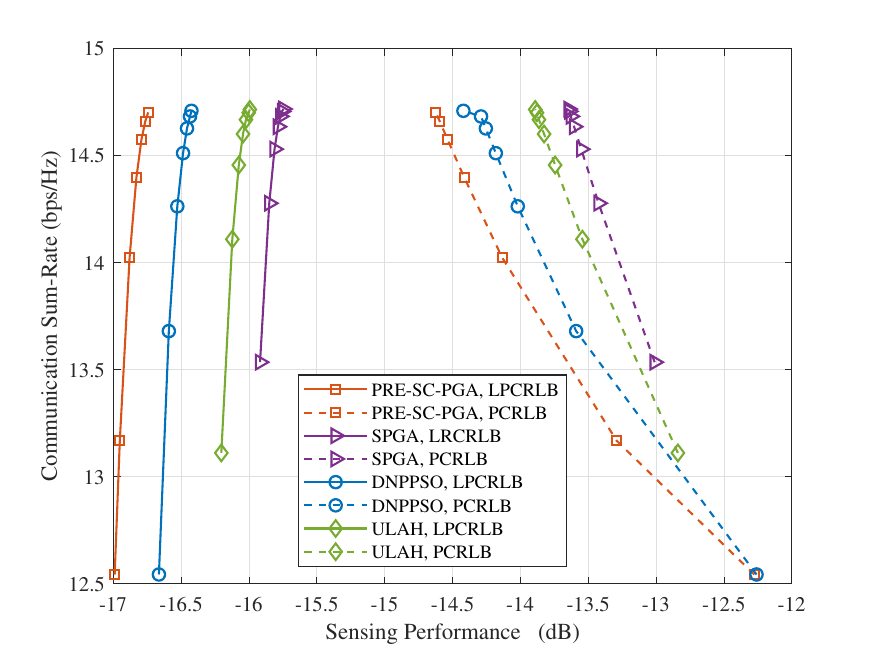}
\caption{\underline{\textbf{This figure is newly added}} The trade-off between communication and angle sensing performance {\color{blue}with $P=1W$, $D_{max}=7\lambda$.}}
\label{fig9}
\end{figure}

\begingroup
\color{blue}
To illustrate the relationship between the LPCRLB and the PCRLB, we take angle estimation accuracy as an example and plot their trade-off curves with respect to the total communication rate, as shown in Fig.  \ref{fig9}.
For the LPCRLB, as the weighting factor increases, the communication performance of all beamforming algorithms improves, whereas the angle estimation accuracy degrades.
Furthermore, by computing the original PCRLB matrix using the optimized parameters obtained for each weighting factor, we obtain the corresponding communication–sensing trade-off curves. 
The resulting estimation accuracy is lower than that of the LPCRLB, confirming that the LPCRLB serves as a lower bound of the PCRLB, which is consistent with the theoretical derivation. 
In addition, a synergistic relationship exists between communication and sensing performance, and under the same communication rate, the proposed PRE-SC-PGA algorithm achieves the best PCRLB performance.
\endgroup

\begingroup
\color{blue}
\subsection{Performance of RPDPSO Algorithm}
The RPDPSO algorithm is specifically designed for maximizing the communication sum-rate under sensing constraints.
The performance obtained with 10 randomly initialized particles over 20 iterations is illustrated in Table III. 
RPDPSO algorithm outperforms the other three algorithms and achieves the best performance.
This superiority stems from the combination of a dynamic granular update mechanism and inertial weights, which allows the algorithm to continuously explore high-reward regions while reducing the risk of premature convergence.
\begin{table}[!t]
\centering
\caption{\underline{\textbf{This table is newly added}}A comparison of the communication sum-rates achieved by the four algorithms in the $(m-1)$-th time slot.}
\footnotesize
\setlength{\tabcolsep}{2pt}
\label{tab4}
\begin{tabular}{|c|c|c|c|c|c|}
\hline
 & ULAH& RPDPSO & DNPPSO & PRE-SC-PGA \\
\hline
Sum-Rate & 14.5143 & 14.7093 & 14.7073 &14.6810 \\
\hline
\end{tabular}
\end{table}

In the $m$-th time slot, the transmit antenna positions optimized from the previous stage are applied and combined with the current channel estimates to redesign the beamforming vectors and power allocation.
As shown in Table IV, the algorithms evaluated show significant differences in the communication sum-rate and perceived LPCRLBs under the given sensing constraints $[2e-4,0.05,1]$.
First, the proposed algorithm continues to achieve substantially higher communication sum-rate compared with the FPA. 
This demonstrates that proactively optimizing the antenna positions to obtain more favorable channel conditions indeed benefits the communication performance in the subsequent time slot.
Moreover, the sensing LPCRLB results show that, while satisfying the sensing constraints, the proposed method even achieves better average estimation accuracy for angle, distance, and velocity than the fixed configuration. These results confirm that the proposed RPDPSO algorithm can simultaneously enhance both communication and sensing performance, thereby validating its effectiveness in dynamic V2I scenarios.

\begin{table}[!t]
\begin{center}
\caption{\underline{\textbf{This table is newly added}}A comparison of the communication sum-rates and sensing LPCRLBs achieved by the four algorithms in the $m$-th time slot}
\footnotesize
\setlength{\tabcolsep}{3pt}
\label{tab6}
\begin{tabular}{|c|c|c|c|c|c|}
\hline
 &ULAH & RPDPSO & DNPPSO & PRE-SC-PGA  \\
 \hline
Sum-Rate & 14.5102 & 14.7047 & 14.7027 & 14.6745 \\
\hline
\multirow{2}{*}{$\mathrm{LPCRLB}_{\theta_k}$} 
& 1.514e-5 & 1.393e-5 & 1.432e-5 & 1.458e-4  \\
\cline{2-6}
~ & 1.375e-5 & 1.230e-5 & 1.270e-5 & 1.303e-5  \\
\hline
\multirow{2}{*}{$\mathrm{LPCRLB}_{d_k}$} 
& 0.0387 & 0.0387& 0.0387 & 0.0387\\
\cline{2-6}
~ & 0.0346 & 0.0343 & 0.0343 & 0.0344 \\
\hline
\multirow{2}{*}{$\mathrm{LPCRLB}_{\nu_k}$} 
& 0.2213 & 0.2216 & 0.2216 & 0.2215\\
\cline{2-6}
~ &0.2303 & 0.2294& 0.2294 & 0.2296  \\
\hline
\end{tabular}
\end{center}
\end{table}

\endgroup

\begingroup
\color{blue}
\section{Conclusion}
This paper presented a comprehensive study on the use of MA technology in dynamic V2I ISAC networks.
By incorporating vehicle motion prediction via an EKF, we jointly optimized the transmit antenna positions, beamforming and power allocation vectors, enabling the system to adapt effectively to rapid channel variations in high-mobility environments.
Two optimization frameworks were developed to address scenarios with and without sensing QoS constraints. 
The PRE-SC-PGA algorithm provides reference and warm-start solutions for unconstrained cases, while the proposed RPDPSO algorithm efficiently handles sensing QoS constraints and non-convexity, achieving gains in both communication sum-rate and sensing estimation accuracy with low computational complexity.

Simulation results demonstrated that MA technology can dynamically adjust the spatial configuration of antennas and beam directions, thereby enhancing link robustness and sensing fidelity in V2I systems. 
These findings highlight the practical relevance of MA-enabled ISAC for real-time decision-making, cooperative perception, and intelligent transportation applications.
In addition, the proposed methodologies validate the feasibility of fast joint optimization under complex non-convex constraints and provide quantitative design insights for future large-scale MA deployments. Potential extensions include multi-RSU cooperative MA optimization, robust designs accounting for CSI and prediction uncertainties, learning-based MA and beam management strategies, and low-complexity multi-objective resource allocation for dense vehicular environments.
\endgroup

\appendices  
\section{Derivation of CRLB}
Equation {\color{blue}(\ref{deqn_ex8a})} in Section {\color{blue}II-B} can be further expressed as
\begingroup
\renewcommand{\arraystretch}{0.8}   
\setlength{\arraycolsep}{3pt} 
\begin{equation}
\scalebox{0.9}{$
\label{deqn_ex57a}
 \mathbf{J} (\mathbf{u} )= \left [ \begin{matrix}
\mathbf{J}  _{11}   & \mathbf{J} _{12} &  \dots& \mathbf{J} _{1K} \\
\mathbf{J}  _{21}   & \mathbf{J}  _{22}& \dots& \mathbf{J}  _{2K}\\
\vdots   &  \vdots&  \ddots & \vdots\\
  \mathbf{J}  _{K1}&  \mathbf{J} _{K2}&  \dots &\mathbf{J}  _{KK}
\end{matrix} \right ], 
$}
\end{equation}
\endgroup
where $\mathbf{J} _{kk}=\boldsymbol{\varpi}^{k\left (H\right ) }\mathbf{E} ^{-1}\boldsymbol{ \varpi}^{k}$ {\color{blue}and $ \boldsymbol{\varpi}^{k}=\left [ \boldsymbol{\varpi}^{1k},\boldsymbol{ \varpi}^{2k},\boldsymbol{ \varpi}^{3k}\right]$. 
Thus, $\mathbf{J} _{kk}$ can be rewritten as
\begin{equation}
\label{deqn_ex58a}
\scalebox{0.85}{$
    \mathbf{J} _{kk}=\left [ \begin{matrix}
\bm{\varpi}^{1k(H)}\mathbf{E}^{-1}\bm{\varpi}^{1k} & \bm{\varpi}^{1k(H)}\mathbf{E}^{-1}\bm{\varpi}^{2k} & \bm{\varpi}^{1k(H)}\mathbf{E}^{-1}\bm{\varpi}^{3k}\\
 \bm{\varpi}^{2k(H)}\mathbf{E}^{-1}\bm{\varpi}^{1k} &  \bm{\varpi}^{2k(H)}\mathbf{E}^{-1}\bm{\varpi}^{2k}&\bm{\varpi}^{2k(H)}\mathbf{E}^{-1}\bm{\varpi}^{3k} \\
 \bm{\varpi}^{3k(H)}\mathbf{E}^{-1}\bm{\varpi}^{1k} & \bm{\varpi}^{3k(H)}\mathbf{E}^{-1}\bm{\varpi}^{2k} &\bm{\varpi}^{3k(H)}\mathbf{E}^{-1}\bm{\varpi}^{3k}
\end{matrix} \right ], $}
\end{equation}
where $\boldsymbol{\varpi}^{ik}= \left[\boldsymbol{\varpi}_{1}^{ik}, \boldsymbol{\varpi}_{2}^{ik}, \cdots, \boldsymbol{\varpi}_{N}^{ik}\right ]^{T}, \boldsymbol{\varpi}_{n}^{ik}=\left[\boldsymbol{\varpi}_{n1}^{ik},\boldsymbol{\varpi}_{n2}^{ik},\notag\right. \\ \left.\cdots,\boldsymbol{\varpi}_{nQ}^{ik}\right]^{T}$.
Substituting (\ref{deqn_ex9a}) into $\boldsymbol{\varpi}_{n}^{ik}$, we can obtain $\boldsymbol{\varpi}^{ik}$.

Next, we take the first motion parameter $\phi_{k}$ as an example to provide the calculation process of $\mathbf{J}_{kk}$. 
Given $\boldsymbol{\varpi}_{nq}^{1k}$ and define $\bm{\mathcal{G}}_{k}=\mathbf{a}(p_{tx},\theta_{k})\mathbf{a}^{H}(p_{tx},\theta_{k})$ and $\bm{\mathcal{K}}_{k}=\mathbf{\Lambda}_{tx}\bm{\mathcal{G}}_{k}\mathbf{\Lambda}_{tx}$, which yields
\begin{equation}
\label{deqn_ex59a}
\begin{aligned}
\boldsymbol{\varpi}_{nq}^{1k(H)}&\sigma_{n}^{-2}\mathbf{I}_{M_{rx}}\boldsymbol{\varpi}_{nq}^{1k}=(\frac{2\pi\gamma_{k}}{\lambda})^{2}\frac{p_{n}}{\eta_{1}T_{e}}\big(\mathrm{tr}(\bm{\Lambda}_{rx}^{2} )\mathbf{w}_{n}^{H}\bm{\mathcal{G}}_{k}\mathbf{w}_{n}\\&+
M_{rx}\mathbf{w}_{n}^{H}\bm{\mathcal{K}}_{k}\mathbf{w}_{n}-2\mathrm{tr}(\bm{\Lambda}_{rx})\mathbf{w}_{n}^{H}\bm{\mathcal{G}}_{k}\bm{\Lambda}_{tx}\mathbf{w}_{n} \big),
\end{aligned}
\end{equation}
where $\mathbf{\Lambda}_{tx}=\mathrm{diag}(\mathbf{p}_{tx})$ and $\mathbf{\Lambda}_{rx}=\mathrm{diag}(\mathbf{p}_{rx})$. Therefore, $\bm{\varpi}^{1k(H)}\mathbf{E}^{-1}\bm{\varpi}^{1k}$ is given by
\begin{equation}
       \label{deqn_ex60a} \bm{\varpi}^{1k(H)}\mathbf{E}^{-1}\bm{\varpi}^{1k}=Q\sum_{n=1}^{N}\boldsymbol{\varpi}_{nq}^{1k(H)}\sigma_{n}^{-2}\mathbf{I}_{M_{rx}}\boldsymbol{\varpi}_{nq}^{1k}.
\end{equation}
Denoting $\mathbf{p}=[p_{1},\cdots,p_{n}]^{T}$, $\mathbf{g}_{kk}^{11}=\left[[\mathbf{g}_{kk}^{11}]_{1},\cdots,[\mathbf{g}_{kk}^{11}]_{N}\right]^{T}$, we have $\bm{\varpi}^{1k(H)}\mathbf{E}^{-1}\bm{\varpi}^{1k}=\mathbf{p}^{T}\mathbf{g}_{kk}^{11}$, where 
\begin{equation}
\label{deqn_ex61a}
\begin{aligned}
\left[\mathbf{g}_{kk}^{11}\right]_{n}&=(\frac{2\pi\gamma_{k}}{\lambda})^{2}\frac{Q}{\eta_{1}T_{e}}\big(\mathrm{tr}(\bm{\Lambda}_{rx}^{2} )\mathbf{w}_{n}^{H}\bm{\mathcal{G}}_{k}\mathbf{w}_{n}+
M_{rx}\mathbf{w}_{n}^{H}\bm{\mathcal{K}}_{k}\mathbf{w}_{n}\\&
-2\mathrm{tr}(\bm{\Lambda}_{rx})\mathbf{w}_{n}^{H}\bm{\mathcal{G}}_{k}\bm{\Lambda}_{tx}\mathbf{w}_{n} \big).
\end{aligned}
\end{equation}
Similarly, other elements of matrix $\mathbf{J}_{kk}$ can also be calculated and we have
\begin{equation}
\label{deqn_ex62a}
\scalebox{0.8}{$
\begin{aligned}
\left[\mathbf{g}_{kk}^{12}\right]_{n}&=-\frac{Q(2\pi\gamma_{k})^{2}n\Delta f}{\lambda\eta_{1}T_{e}}(\mathrm{Tr}(\mathbf{\Lambda}_{rx})\mathbf{w}_{n}^{H}\bm{\mathcal{G}}_{k}\mathbf{w}_{n}-M_{rx}\mathbf{w}_{n}^{H}\bm{\Lambda}_{tx}\bm{\mathcal{G}}_{k}\mathbf{w}_{n}),\\
\left[\mathbf{g}_{kk}^{13}\right]_{n}&=\frac{Q(Q-1)(2\pi\gamma_{k})^{2}T_{s}}{2\lambda\eta_{1}T_{e}}(\mathrm{Tr}(\mathbf{\Lambda}_{rx})\mathbf{w}_{n}^{H}\bm{\mathcal{G}}_{k}\mathbf{w}_{n}-M_{rx}\mathbf{w}_{n}^{H}\bm{\Lambda}_{tx}\bm{\mathcal{G}}_{k}\mathbf{w}_{n}),\\
\left[\mathbf{g}_{kk}^{22}\right]_{n}&=\frac{Q(2\pi\gamma_{k}n\Delta f)^{2}}{\eta_{1}T_{e}}M_{rx}\mathbf{w}_{n}^{H}\bm{\mathcal{G}}_{k}\mathbf{w}_{n},\\
\left[\mathbf{g}_{kk}^{33}\right]_{n}&=\frac{2Q(Q-1)(2Q-1)(\pi\gamma_{k}T_{s})^{2}}{3\eta_{1}T_{e}}M_{rx}\mathbf{w}_{n}^{H}\bm{\mathcal{G}}_{k}\mathbf{w}_{n},\\
\left[\mathbf{g}_{kk}^{23}\right]_{n}&=-\frac{2Q(Q-1)(\pi\gamma_{k})^{2}n\Delta fT_{s}}{\eta_{1}T_{e}}M_{rx}\mathbf{w}_{n}^{H}\bm{\mathcal{G}}_{k}\mathbf{w}_{n}.
\end{aligned}$}
\end{equation}}

\section{convergence Analysis of Beamforming Algorithm}
Ignoring the rank-one constraint, the optimization problem {\color{blue}$(\mathrm{P1.2})$} can be reformulated as
{\color{blue}
\begin{equation}
\label{deqn_ex63a}
\scalebox{0.9}{$
\begin{aligned}
\max_{\mathbf{W},\boldsymbol{\kappa}, \boldsymbol{\epsilon},\boldsymbol{\varepsilon}}\,\,\,&\mathcal{F}_{11}(\boldsymbol{\kappa}, \boldsymbol{\epsilon},\boldsymbol{\varepsilon})+\mathcal{F}_{21}(\mathbf{W},\boldsymbol{\kappa}, \boldsymbol{\epsilon},\boldsymbol{\varepsilon})\\
\mathrm{s.t.}\,\,& \rm{(\ref{deqn_ex41A})},  \rm{(\ref{deqn_ex41C})-(\ref{deqn_ex41F})}.
\end{aligned}$}
\end{equation}}
All constraints in the above optimization problem are convex and {\color{blue}$\mathcal{F}_{11}(\boldsymbol{\kappa}, \boldsymbol{\epsilon},\boldsymbol{\varepsilon})$} in the objective function is non-concave. 
In order to solve this non-convex problem, we use successive convex approximation to approximate it as a series of convex problems and then solve them iteratively. 
Assuming that the local points of each iteration are ${\color{blue}\mathbf{W}_{n}^{ir}}, \kappa_{k}^{ir}, \epsilon_{k}^{ir}$ and $\varepsilon_{k}^{ir}$, based on the first-order Taylor expansion, the global linear lower bound function of {\color{blue}$\mathcal{F}_{11}(\boldsymbol{\kappa}, \boldsymbol{\epsilon},\boldsymbol{\varepsilon})$} at the local points can be expressed as {\color{blue}$\breve{\mathcal{F}}_{11}^{(ir)}(\boldsymbol{\kappa}, \boldsymbol{\epsilon},\boldsymbol{\varepsilon})$.}

By replacing {\color{blue}$\hat{\mathcal{F}}_{11}( \boldsymbol{\kappa}, \boldsymbol{\epsilon},\boldsymbol{\varepsilon})$} with {\color{blue}$\breve{\mathcal{F}}_{11}^{(ir)}(\boldsymbol{\kappa}, \boldsymbol{\epsilon},\boldsymbol{\varepsilon})$}, the objective function in inner iteration $ir$ can be approximated as {\color{blue}$\breve{\mathcal{F}}_{11}^{(ir)}(\boldsymbol{\kappa}, \boldsymbol{\epsilon},\boldsymbol{\varepsilon})+\hat{\mathcal{F}}_{21}(\mathbf{W},\boldsymbol{\kappa}, \boldsymbol{\epsilon},\boldsymbol{\varepsilon})$.} 
The iterative process of local points follows
{\color{blue}
\begin{equation}
\label{deqn_ex64a}
\scalebox{0.9}{$
\begin{aligned}
\hat{\mathcal{F}}_{11}&(\boldsymbol{\kappa}^{ir+1}, \boldsymbol{\epsilon}^{ir+1},\boldsymbol{\varepsilon}^{ir+1}) + \hat{\mathcal{F}}_{21}(\mathbf{W}^{ir+1}, \boldsymbol{\kappa}^{ir+1}, \boldsymbol{\epsilon}^{ir+1},\boldsymbol{\varepsilon}^{ir+1})\\&\ge\breve{\mathcal{F}}_{11}^{(ir)}(\boldsymbol{\kappa}^{ir+1}, \boldsymbol{\epsilon}^{ir+1},\boldsymbol{\varepsilon}^{ir+1})+\hat{\mathcal{F}}_{21}(\mathbf{W}^{ir+1},\boldsymbol{\kappa}^{ir+1}, \boldsymbol{\epsilon}^{ir+1},\boldsymbol{\varepsilon}^{ir+1})\\&\ge\breve{\mathcal{F}}_{11}^{(ir)}(\boldsymbol{\kappa}^{ir}, \boldsymbol{\epsilon}^{ir},\boldsymbol{\varepsilon}^{ir})+\hat{\mathcal{F}}_{21}(\mathbf{W}^{ir},\boldsymbol{\kappa}^{ir}, \boldsymbol{\epsilon}^{ir},\boldsymbol{\varepsilon}^{ir})\\&=\hat{\mathcal{F}}_{11}( \boldsymbol{\kappa}^{ir}, \boldsymbol{\epsilon}^{ir},\boldsymbol{\varepsilon}^{ir})+\hat{\mathcal{F}}_{21}(\mathbf{W}^{ir}, \boldsymbol{\kappa}^{ir}, \boldsymbol{\epsilon}^{ir},\boldsymbol{\varepsilon}^{ir}).
\end{aligned}$}
\end{equation}}
Obviously, the objective sequence constructed in the optimization problem $\mathrm{(P2.1)}$ is bounded and non-decreasing, which ensures convergence to a stationary solution.

\end{document}